\documentclass[sigplan, nonacm, screen, 10pt]{acmart}
\usepackage{mathrsfs}
\usepackage{algorithm}
\usepackage[noend]{algpseudocode}
\usepackage{amsmath}
\usepackage{enumitem}
\usepackage{booktabs}
\usepackage{subcaption}

\AtBeginDocument{%
  }

\setcopyright{acmlicensed}
\copyrightyear{2018}
\acmYear{2018}
\acmDOI{XXXXXXX.XXXXXXX}
\begin{document}

\title[Orchestrate Multimodal Data with Batch Post-Balancing to Accelerate MLLM Training]{OrchMLLM: Orchestrate Multimodal Data with Batch Post-Balancing to Accelerate Multimodal Large Language Model Training}

\pagestyle{plain}

\author{
\rm{Bangjun Xiao$^{\text{1,2}, *}$ \enskip
    Yijie Zheng$^{\text{1}, *}$ \enskip
    Lei Shi$^{\text{1}, *}$ \enskip
    Xiaoyang Li$^{\text{1}}$  \enskip
    Faming Wu$^{\text{1}}$  \enskip}
\\
\rm{    
    Tianyu Li$^{\text{1}}$ \enskip
    Xuefeng Xiao$^{\text{1}}$ \enskip
    Yang Zhang$^{\text{1}}$ \enskip
    Yuxuan Wang$^{\text{1}}$  \enskip
    Shouda Liu$^{\text{1}}$ \enskip}
\\
{$^{\text{1}}$ByteDance Seed\enskip $^{\text{2}}$Peking University}
}

\renewcommand{\shortauthors}{Xiao et al.}

\begin{abstract}
{\let\thefootnote\relax\footnote{{$^*$Equal contribution.}}}
Multimodal large language models (MLLMs), such as GPT-4o, are garnering significant attention. During the exploration of MLLM training, we identified Modality Composition Incoherence, a phenomenon that the proportion of the same modality varies dramatically across different examples. It exacerbates the challenges of addressing mini-batch imbalances, which lead to uneven GPU utilization between Data Parallel (DP) instances and severely degrades the efficiency and scalability of MLLM training, ultimately affecting training speed and hindering further research on MLLMs.

To address this challenge, we introduce OrchMLLM, an efficient and adaptive framework designed to mitigate the inefficiencies in MLLM training caused by mini-batch imbalances. First, we propose Batch Post-Balancing Dispatcher, a technique that efficiently eliminates mini-batch imbalances of sequential data. Additionally, we integrate MLLM Global Orchestrator into the training framework to orchestrate multimodal data and tackle the issues arising from Modality Composition Incoherence. We evaluate OrchMLLM across various MLLM sizes, demonstrating its efficiency and scalability. Experimental results reveal that OrchMLLM achieves a Model FLOPs Utilization (MFU) of $41.6\%$ when training an 84B MLLM with three modalities on $2560$ H100 GPUs, outperforming Megatron-LM by up to $3.1\times$ in throughput.


\end{abstract}

\maketitle

\section{Introduction}
The success of large language models (LLMs) is sparking a revolution in AI applications~\cite{chatgpt}. Given the diverse modalities that constitute information in the real world, there is a growing expectation for unified models capable of processing multimodal information. To bridge the gap of LLMs that only focus on text, emerging multimodal large language models (MLLMs) are integrating various media types, including text, images, and audio. 
Initially, MLLMs combine the text modality with another modality, including GPT-4V~\cite{gpt4-v}, Gemini~\cite{gemini}, and CogVLM~\cite{wang2024cogvlmvisualexpertpretrained} for the visual modality, as well as Qwen-Audio~\cite{chu2023qwenaudioadvancinguniversalaudio} and Seed-ASR~\cite{bai2024seedasrunderstandingdiversespeech} for the auditory modality. Recently, those MLLMs that integrate more modalities into a single MLLM, i.e. omni models, are continuously attracting attention, such as GPT-4o~\cite{gpt4-o}, Qwen2.5-Omni~\cite{xu2025qwen25omnitechnicalreport} and MiniCPM-o~\cite{yao2024minicpm}, etc. These advancements indicate that there is a continuous effort to explore the potential of MLLMs.

\begin{figure}[t]
  \centering
  \includegraphics[width=\linewidth]{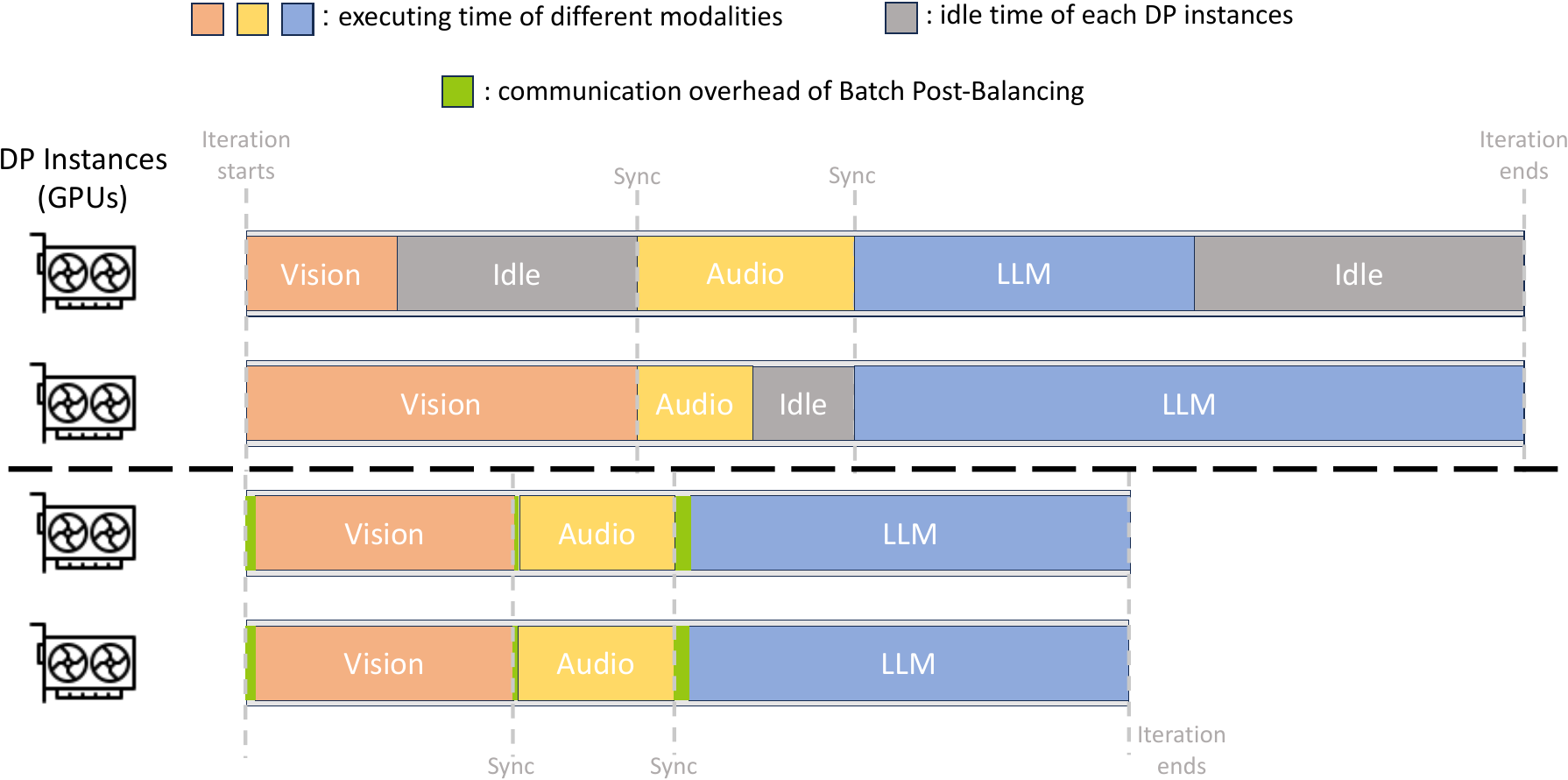}
  \vspace{-0.15in}
  \caption{A comparative illustration showcasing the effectiveness of OrchMLLM. }
  \label{fig:dia}
  \vspace{-0.25in}
\end{figure}

Owing to the scaling law, considerable resources are dedicated to training MLLMs with billions of parameters and trillions of tokens, which is extremely expensive and time-consuming~\cite{team2024chameleon, treport}. 
Currently, several existing frameworks such as Megatron-LM~\cite{shoeybi2019megatron} and DeepSpeed~\cite{10.1145/3394486.3406703} can be adapted to accelerate the training speed of MLLMs.
However, when training with sequential data, the varying sequence lengths and the randomness in batching~\cite{Robbins1951ASA} introduce imbalances in mini-batches across different data parallelism (DP) instances. This imbalance causes uneven GPU utilization, which consequently degrades the efficiency and scalability of MLLM training. 
Furthermore, independent execution of encoders indicates multiple mini-batches during different phases and requires the elimination of imbalances during each phase, otherwise, mini-batch imbalances of arbitrary modalities can significantly reduce training efficiency. The challenge to achieve this objective is raised by a phenomenon in multimodal data, called \textbf{Modality Composition Incoherence}, which refers to the dramatic variation in the proportion of the same modality across different examples. 
Modality Composition Incoherence makes it difficult to resolve mini-batches imbalances across all phases through performing balancing operations only on mini-batches of examples. Existing methods, collectively referred to as Batch Pre-Balancing methods in this paper, operate only on mini-batches of examples at the beginning of a training iteration and can focus on the imbalance only in a single phase, inevitably failing to achieve fully accelerate MLLM training.

To address this challenge, we present OrchMLLM, an efficient and adaptive framework designed to comprehensively resolve mini-batch imbalances and accelerate MLLM training, as illustrated in Figure~\ref{fig:dia}.
The core insight of OrchMLLM is that rearranging mini-batches across DP instances does not affect training results (i.e., the rearrangement is consequence-invariant) and that a proper rearrangement of mini-batches can eliminate the imbalance and further balance GPU utilization. Building on this observation, Batch Post-Balancing Dispatcher is proposed to achieve balance among mini-batches of a single modality after mini-batches have been decided, breaking the complex problem into sub-problems for each phase. First, we formulate the problem and propose several Batch Post-Balancing algorithms, which prepare the dispatcher to determine the appropriate rearrangement for mini-batches in different scenarios. Furthermore, we devise a Node-wise All-to-All Communicator to implement the practical rearrangement of mini-batches. Specifically, All-to-All Batch Communicator rearranges mini-batches with lightweight communication overhead and memory occupancy, and Node-wise Rearrange Algorithm further reduces the communication overhead by leveraging heterogeneous bandwidths between intra-node and inter-node instances. Lastly, the custom-designed MLLM Global Orchestrator is integrated into the MLLM training workflow, comprehensively resolving mini-batch imbalances during each phase and significantly enhancing training efficiency.

 
In summary, we make the following contributions:

 \begin{itemize}[leftmargin=10pt]
\item We present OrchMLLM, an efficient and adaptive framework that comprehensively addresses mini-batch imbalances and accelerates the training of MLLMs. OrchMLLM is also applicable to all large-scale distributed training with sequential data, regardless of the model architectures in training, without requiring much operator code refactoring.

\item We propose the Batch Post-Balancing Dispatcher, a technique that efficiently eliminates mini-batch imbalances in sequential data. Additionally, we integrate the MLLM Global Orchestrator into OrchMLLM to orchestrate multimodal data and resolve the challenges posed by Modality Composition Incoherence.

\item We implement OrchMLLM and conduct experiments on the cluster with 2560 H100 GPUs. The results show that OrchMLLM achieves $41.6\%$ MFU when training an 84B MLLM with both visual and auditory modalities, outperforming Megatron-LM by up to 3.1× in throughput.

 \end{itemize}


\section{Background}
In this section, we provide background knowledge for multimodal large language model (MLLM) training.

\subsection{MLLM training}
\label{sec:train}
A typical MLLM is comprised of three types of modules~\cite{yin2024surveymultimodallargelanguage}: a pretrained LLM backbone, several single-modality encoders (e.g. ViT\cite{dosovitskiy2020image} for vision and Whisper\cite{whisper} for audio), and respective connectors for bridging between the encoders and the LLM backbone. A common and effective implementation of connectors is to transform the output features of encoders into tokens and concatenate all these tokens before sending them to the LLM backbone. In MLLM, we collectively refer to the LLM backbone and the encoders (along with their corresponding connectors) as \textit{submodules}.

When training MLLMs, the training dataset is composed of multimodal data examples. In the forward pass, metadata from different modalities, including images, audios, etc., will be organized as sequential data (e.g. restructuring images into sequences of patches), processed by the corresponding encoders and transformed into sequences of tokens by connectors. These tokens belong to the unified embedding space of the LLM, and the encoded sequences comprised of them are termed \textit{subsequences} within the whole sequence. After encoding, an example's subsequences of different modalities are interleaved according to the order predefined by the example or certain templates, and the entire sequence is processed by the LLM backbone. The execution of each submodule is termed a \textit{phase} of an iteration.

\subsection{Data Parallelism}
\label{sec:dp}
 Data parallelism (DP) is a technique used to scale training across multiple devices by distributing data. In classic DP~\cite{dp1, Dean2012LargeSD}, all model parameters and optimizer states are replicated on each device. At each training step, a global batch is divided into several subsets (i.e. \textit{mini-batches}) across all DP instances. Each instance executes the forward and backward propagation on a different mini-batch and reduces gradients across instances to update the model locally. 
 
For models with a large number of parameters, a complete replica of the model cannot fit in the device memory. Therefore, some studies have proposed variants of DP that proportionally reduce the memory footprint. ZeRO (Zero Redundancy Optimizer)~\cite{rajbhandari2020zero} is a memory-efficient variant of DP where the model states are partitioned across all devices instead of being replicated. These states are reconstructed using gather-based communication collectives on-the-fly during training. 
Fully sharded data parallelism (FSDP)~\cite{zhao2023pytorch}, an efficient and user-friendly implementation of ZeRO, overlaps the communication of shards with computation on the critical path, thus mitigating the impact on training efficiency. By reducing the memory requirement for each device, these variants enable the scale-up of models that could not previously be trained due to memory limitations. This opens up opportunities for researchers to explore larger and more powerful models, including MLLMs.

\subsection{Imbalance in Mini-batches}
\label{sec:sd}

\begin{figure}[t]
  \centering
  \includegraphics[width=\linewidth]{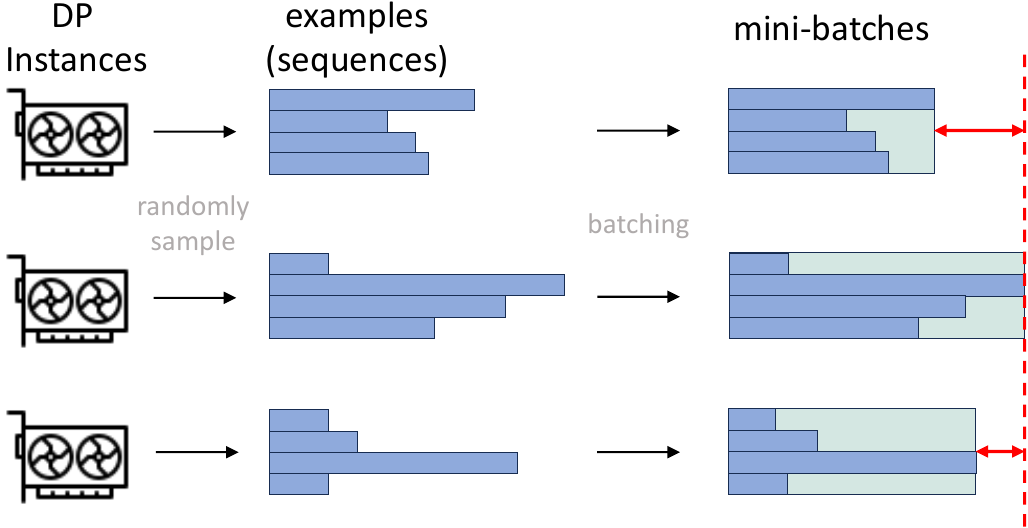}
  \caption{The diagram for the imbalance in mini-batches in the training of sequential data, using the padding batching method as an example. }
  \label{fig:imbalance}
  \vspace{-0.15in}
\end{figure}
Currently, many deep learning tasks organize input data as sequential data for processing (including texts, audios, and patches from images), especially with the rise of Transformer architecture~\cite{vaswani2017attention}.
However, in terms of the training of these models, the sequential data in training datasets exhibit large variance in terms of sequence lengths (e.g. ranging from 10 to 40k, and even larger, in production datasets).
Compared with the fixed-sized input data, a significant property of a batch is not the batch size anymore, but the token count, which equals the post-padding sequence length multiplied by the batch size when padding is employed, or otherwise the sum of the sequence lengths.
Given the fact that the algorithm of Batch Gradient Descent adheres to the principle of batching randomness~\cite{Robbins1951ASA}, the token count can be regarded as a random variable. Therefore, the token counts of several randomly selected batches can significantly deviate from each other~\cite{sd}. In the context of DP, where each DP instance randomly samples mini-batches from the dataset, this substantial variance of token counts remains across the mini-batches on different instances during the training with sequential data, which we refer to as \textit{imbalance in mini-batches}.

Because the token count is strongly correlated to the computation cost and memory occupancy for a batch~\cite{vaswani2017attention}, the imbalance in mini-batches leads to many problems when training with sequential data. 
First, during the synchronized communication between instances (required for any variant of DP~\cite{Dean2012LargeSD, rajbhandari2020zero,zhao2023pytorch}), an instance that processes a mini-batch with a small token count is forced to wait for other instances to perform synchronized operations.
Additionally, the memory occupied by the activations of the training data is proportional to the token count of a mini-batch. Assuming that the device memory on DP instances is finite and homogeneous (which is typical in real-world hardware), to avoid the out-of-memory (OOM) error, the batch size must be determined by the maximum token count of all mini-batches. This restricted batch size leads to low memory utilization during training, resulting in most mini-batches not being sufficiently vectorized.
To sum up, aforementioned problems lead to lower GPU utilization, significantly slowing down the training with sequential data. 

\section{Motivation and Challenges}
In the workflow of MLLM training (\S\ref{sec:train}), a sampled mini-batch encompasses several multimodal examples, which contain metadata of different modalities. In this mini-batch, all metadata of the same modality will be reorganized into a new mini-batch to facilitate encoding in the corresponding phase. Finally, the encoded features (i.e. subsequences) will be respectively interleaved into the complete sequences of these examples according to the predefined orders and generate a new mini-batch for the phase of LLM backbone.
Because training data from different modalities are usually organized as sequential data, imbalance in mini-batches (\S\ref{sec:sd}) cannot be ignored in each phase during MLLM training. Furthermore, because the phases of encoders inevitably occupy a significant portion of the execution time and memory throughout each iteration~\cite{feng2024optimus}, it is crucial to address the imbalance in mini-batches in each phase to fully unleash the potential of accelerators. However, in the presence of the phenomenon called \textbf{Modality Composition Incoherence}, achieving this objective becomes quite challenging.

\subsection{Modality Composition Incoherence}
\label{sec:mod}
To build a unified multimodal model, especially an omni model, it is essential for MLLMs not only to comprehend different modalities but also to be capable of performing a variety of tasks. These tasks include those inherent to each modality as well as complex tasks arising from the combination of multiple modalities. Consequently, the dataset used for training MLLMs, especially during the instruction tuning stage, usually contains a wide variety of tasks to enrich the MLLM's capability.
 The diversity of tasks raises an issue regarding the composition of multimodal data, \textbf{Modality Composition Incoherence}, in MLLM training. Intuitively, examples for the same task may exhibit certain common characteristics in the composition of multimodal data. For instance, datasets for automatic speech recognition (ASR) comprise paired data, including auditory data and text data that represent the recognition results. The sequence length of auditory data has a significant positive correlation with the sequence length of text data, as longer speech is generally transcribed into longer text. 

\begin{figure}[t]   
    \centering
    \begin{subfigure}{0.48\linewidth} 
        \centering
        \includegraphics[width=\linewidth]{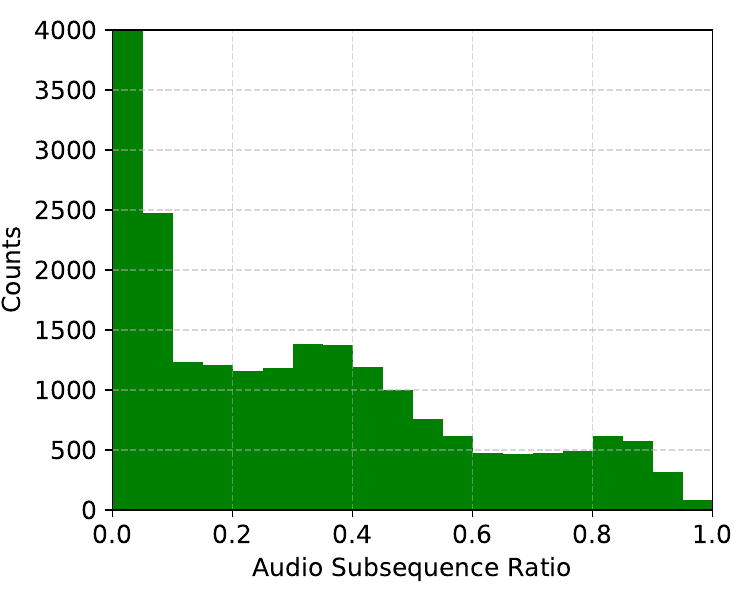} 
        \caption{Audio}
        \label{fig:subfig1}
    \end{subfigure}
    \hfill 
    \begin{subfigure}{0.48\linewidth} 
        \centering
        \includegraphics[width=\linewidth]{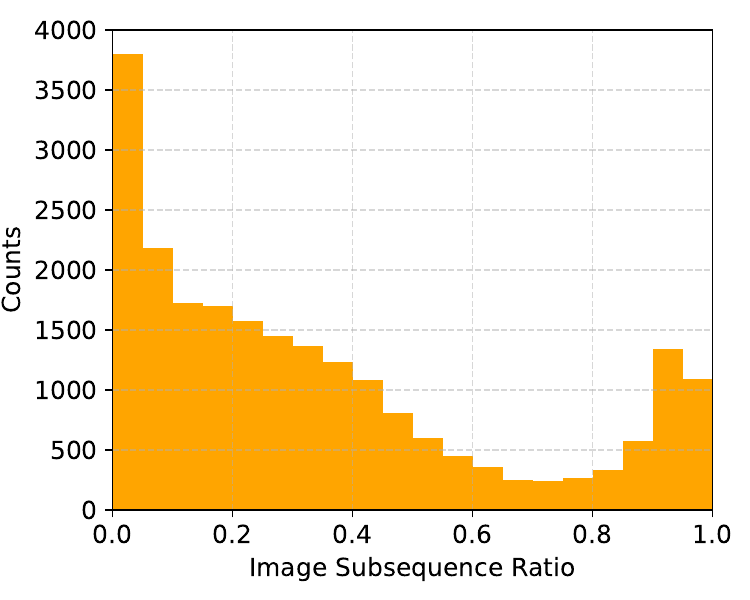} 
        \caption{Image}
        \label{fig:subfig2}
    \end{subfigure}
     \vspace{-0.1in}
    \caption{Modality Composition Incoherence in MLLM training, where the statistics data comes from sampled production datasets. Both of the ratios bear substantial variance.}
 
\vspace{-0.2in}
    \label{fig:comp}
\end{figure}

 However, the composition of multimodal data varies dramatically between different tasks. For the spoken question answering task, there is no direct correlation between the sequence lengths of auditory data and text since a long question in auditory data may receive just a 'yes' or 'no' response. Moreover, the incoherence of composition becomes more pronounced across tasks involving different modalities. The auditory data within an example for the image caption task is evidently absent, and conversely, visual data is missing for the auditory tasks.
Owing to the strict proportional relationship between the sequence length of metadata and the lengths of subsequences, Modality Composition Incoherence can be intuitively characterized by the proportion of the subsequence lengths within the complete sequence. As shown in Figure~\ref{fig:comp}, the proportions of subsequence lengths from visual and auditory modalities both exhibit substantial variance.

As we have analyzed, the decision on a mini-batch of examples during MLLM training is accompanied by different mini-batches in different phases, and the incoherence indicates no fixed composition among these mini-batches. Because imbalance in mini-batches leads to imbalance in computing duration and memory occupancy (\S\ref{sec:sd}), a DP instance may play different roles during different phases, i.e. may be idle during one phase and yet transform into a straggler in another, as shown in Figure~\ref{fig:dia}. Therefore, carefully selecting examples to form mini-batches with the goal of achieving balance in a particular phase cannot eliminate imbalances in all phases, making it quite challenging to fully accelerate MLLM training. The essence of the challenge lies in the fact that, the problem of achieving balance across all phases by balancing the mini-batches of examples becomes a multi-objective optimization problem, which is much more complex than the case with a single modality.

\subsection{Limitation of Existing Methods}
\label{sec:existing}
Several methods have been proposed to address the imbalance in mini-batches in pursuit of accelerating DP training. The key strategy is to form all mini-batches with the same token count~\cite{ye2022dbsdynamicbatchsize}. A straightforward solution is to adopt a dynamic batch size, replacing the fixed batch size with an upper bound for the token counts of mini-batches. Improvements on this method include refining the batching strategy with more complex algorithms, such as using several buckets to store the data and batching the data once a bucket is filled. Although these improvements enhance the balancing effectiveness, they also compromise on the principle of batching randomness. Moreover, these methods, which target the training of a single modality, cannot comprehensively resolve the issue of imbalance in MLLM training.

In additional, DistTrain~\cite{zhang2024disttrainaddressingmodeldata} targets MLLM training and mentions the issue of data heterogeneity, which is similar to Modality Composition Incoherence. This work attempts to address this challenge during data preprocessing by fixing the sequence length in the phase of the LLM backbone and then globally balancing the image input, which is the only modality except textual data in the context of this work.
In this way, this method simplifies the problem by addressing the imbalance within a single modality. However, it fails to solve the problem of training MLLMs with three or more modalities, like these omni-models. In addition, fixing the sequence length in a certain phase cannot always be reasonable, especially during the instruction tuning stage.

In summary, since all these methods perform the balancing when generating mini-batches of examples and before the forward pass, we collectively refer to these approaches as Pre-Balancing methods. This implies that they need to address this multi-objective optimization problem all at once, as analyzed in \S\ref{sec:mod}, which is quite challenging. Consequently, this approach cannot comprehensively resolve the imbalance in mini-batches in MLLM training and achieve the theoretical upper limit of efficiency.

\vspace{-0.05in}
\subsection{Opportunity}
\label{sec:opp}
Since addressing the batch balancing problem all at once is extremely challenging, an alternative approach is to decompose it into multiple single-objective optimization problems. However, given that the mini-batches in all phases are determined simultaneously, it is necessary to carry out the balancing algorithm after the mini-batches of examples have been decided.

Fortunately, we observe that after each DP instance randomly samples examples from the dataset, any permutation or rearrangement across DP instances of these examples will not affect the final gradients used to update the model parameters. 
Across DP instances, the model parameters used for computation in the forward pass remain consistent, hence the computational result for each example is independent of the instance it resides in. Moreover, since all operations for the loss and gradients involve all-reduce, which satisfies the commutative and associative laws, any permutation or rearrangement within all the examples across DP instances is consequence-invariant. 

This observation makes it possible to eliminate the imbalance in mini-batches after they have been decided. We refer to this approach as post-balancing. By performing post-balancing before each phase in MLLM training, we comprehensively address the inefficiency during training.
\section{OrchMLLM Overview}
\begin{figure}[t]
  \centering
  \includegraphics[width=\linewidth]{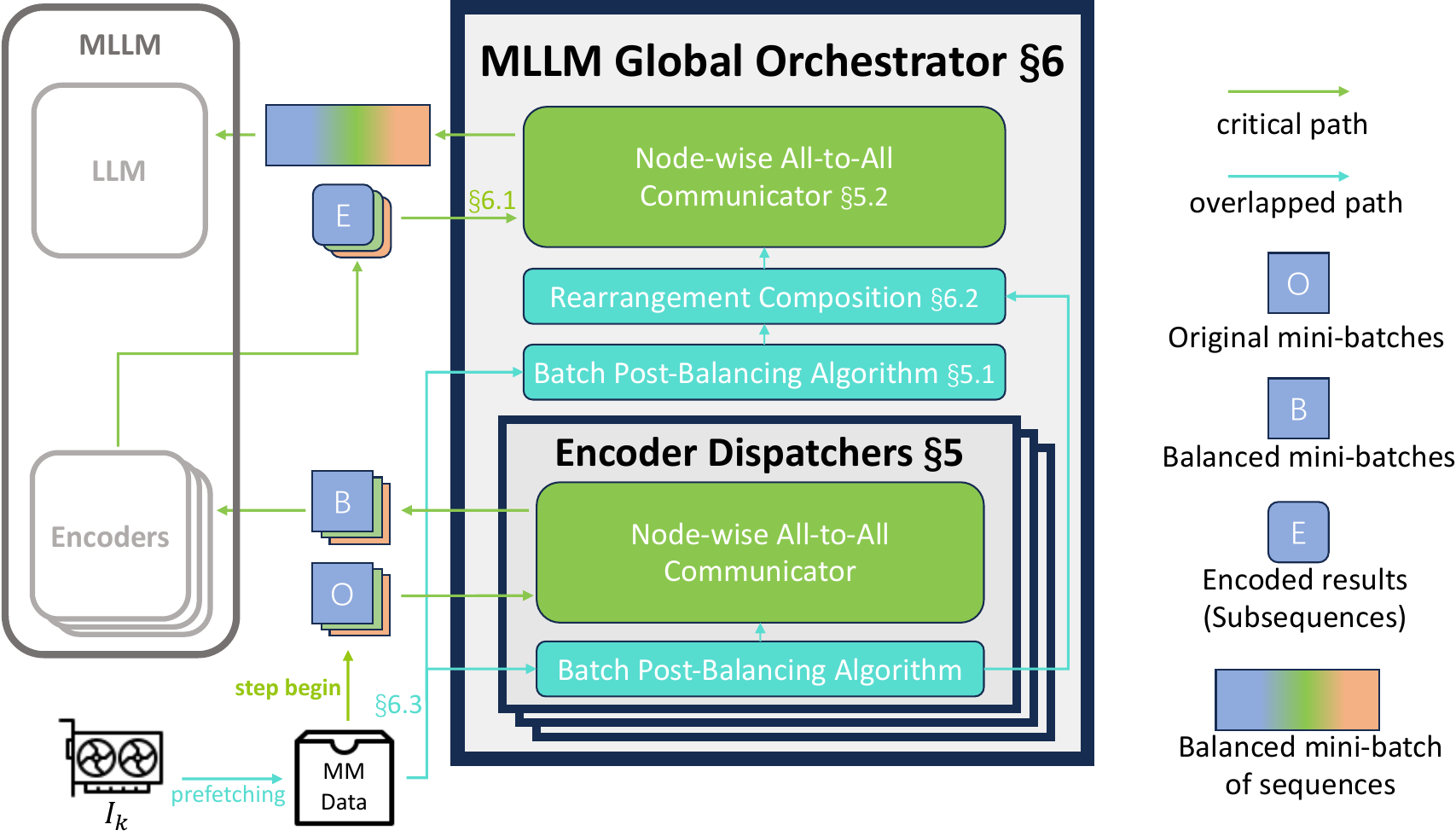}
   \vspace{-0.3in}
  \caption{The system overview of OrchMLLM. }
  \label{fig:overview}
   \vspace{-0.15in}
\end{figure}

We present OrchMLLM, a distributed training framework optimized for MLLM training. The core insight of OrchMLLM is based on the opportunity mentioned above. Firstly, we propose Batch Post-Balancing Dispatcher, a technique which can efficiently eliminate the imbalance in mini-batches of sequential data. Further, we integrate MLLM Global Orchestrator into the framework of MLLM training to orchestrate the multimodal data. By eliminating the imbalance in mini-batches for all phases of MLLM training, our framework can comprehensively solve the inefficiency caused by multimodal data. Therefore, OrchMLLM boosts the GPU utilization and accelerate the training speed, achieving higher efficiency and scalability of MLLM training. A brief overview of the diagram is also shown in Figure~\ref{fig:overview}.

\medskip\noindent \textbf{Batch Post-Balancing Dispatcher.} To eliminate the imbalance in mini-batches of single modality, the problem how to perform batch rearrangement to achieve optimal balance is formulated. Moreover, considering the diverse functional relationships between mini-batches and resource consumption, several balancing algorithms are devised to adapt to different scenarios. In addition, the dispatcher uses the Node-wise All-to-All Communicator to reduce communication overhead and memory occupancy during rearrangement. This communicator not only implements the batch rearrangement efficiently, but also takes into consideration the heterogeneous bandwidths between intra-node and inter-node instances within a cluster, further reducing communication overhead.

\medskip\noindent  \textbf{MLLM Global Orchestrator} MLLM Global Orchestrator is custom-designed for the workflow of MLLM training. In the forward pass, it executes the batch dispatcher of each encoder independently. Owing to the dependencies between the LLM backbone and encoders, MLLM Global Orchestrator will perform the Batch Post-Balancing Algorithm globally and rearrange the data of all modalities. Besides, MLLM Global Orchestrator performs Rearrangement Composition to reduce the rearrangement overhead between encoders and the LLM backbone and overlaps the computation of dispatchers on the non-critical path.


\section{Batch Post-Balancing Dispatcher}
In this part, we will concentrate on the imbalance across DP instances on the dataset which only involves sequential data of single-modality. 
\label{sec:dispatcher}
\subsection{Batch Post-Balancing Algorithms}
\label{sec:alg}
 Because any permutation or rearrangement among the examples of all mini-batches across DP instances will not affect the training results (\S\ref{sec:opp}), we devise algorithms to find the optimal rearrangement to address the imbalance.
 
\medskip\noindent\textbf{Problem formulation.}
Assume there are $d$ DP instances involved in the training, $\mathcal{I}_0, I_1, \dots, I_{d-1}$. Each DP instance randomly samples a mini-batch from the dataset. For $0\leq i < d$, the mini-batch of instance $I_i$ can be regarded as a set comprised of several examples. We denote this set by $S_i$ and the number of examples in this set by $b_i$. Each example is sequential data with length $l_{i, j}$ ($0 \leq i < d,~ 0 \leq j < b_i$). Therefore, the batch length of the mini-batch for instance $I_i$, denoted by $L_i$, is given by: 
\begin{equation} \label{equ:Li}
    L_i=\left\{
\begin{aligned}
\max_{0\leq j < b_i} b_i l_{i,j},~~~ & \text{if padding},\\
\sum_{j=0}^{b_i - 1} l_{i,j},~~~~~~ & \text{otherwise}. \\
\end{aligned}
\right.
\end{equation}
Next, we rearrange all the examples across the $d$ instances into $d$ new mini-batches. In other words, a rearrangement $\Pi$ is defined by the following mapping:
\begin{align*}
    \Pi:\quad \mathcal{S} &\to \mathcal{S} \\
         M &\mapsto M',
\end{align*}
where $M, M' \in \mathcal{S}$ with $\mathcal{S}$ representing the collection of all matrices with dimension $d \times \sum_{i=0}^{d-1} b_i $, and the $(i,j)$-entry of $M$ is rearranged as the $(i',j')$-entry of $M'$. This rearrangement maps example $j$ of the original mini-batch $i$ into the $j'$-th example of the $i'$-th new mini-batch, $S_i'(\Pi)$. For a given rearrangement $\Pi$, we denote the batchsize of $S_i'(\Pi)$ as $b_i'(\Pi)$, and its new batch length is calculated by \eqref{equ:Li} as $L_i'(\Pi)$. Our goal is to eliminate the imbalance in these new mini-batches by finding the optimal rearrangement $\Pi$ that minimizes the following minimax rule:
\begin{equation*}
 \text{Objective:} \quad  \underset{\Pi}{\text{minimize}} ~ \max_{0\leq i < d} f(S_i'(\Pi)),
\end{equation*}
where $f$ is the function which represents the value of computational cost (basically proportional to memory usage) in the training for a given set of examples $S_i'(\Pi)~ (0 \leq i < d)$. The practical significance of this objective is that by minimizing the maximum of the computational costs across the instances, we could maximize the GPU utilization of the whole system, further accelerating the training speed.

As we can see, this problem is a load balancing problem, which can be reduced to the Subset Sum Problem (SSP) and is apparently an NP-complete problem. Therefore, we implement corresponding approximation algorithms based on the organization of sequential data (whether padding or not) and model architecture (determines the function $f$), in order to complete the solution within a polynomial time.
Specific to the Transformer architecture commonly used in MLLMs, the function $f$ derives from the computation complexity of Transformer~\cite{mqa, gqa} and is given by:
\begin{equation} \label{equ:func}
    f(S_i):=\left\{
\begin{aligned}
\alpha L_i + \frac{1}{b_i} \beta L_i^2,~~~~~~~ & \text{if padding},\\
\alpha L_i + \beta\sum_{j=0}^{b_i - 1} l_{i,j}^2,~~~ & \text{otherwise}, \\
\end{aligned}
\right.
\end{equation}
where $\alpha$ and $\beta$ are the constant values determined by the model architecture, and $l_{i,j} (0 \leq j < b_i -1)$ are the sequence lengths of the examples in the set $S_i$. Typically, there is an assumption that $\beta \ll \alpha$, so that the function $f$ in~\eqref{equ:func} can be approximated by
$f(S_i) := \alpha L_i$, and the objective is:
\begin{equation*}
 \text{Objective:} \quad  \underset{\Pi}{\text{minimize}} ~ \max_{0\leq i < d} L_i'(\Pi).
\end{equation*}

\begin{algorithm}[t!]
\caption{Post-Balancing Algorithm without Paddings}\label{alg:rmpad}
    \begin{footnotesize}
    \begin{algorithmic}[1]
        \Require count of DP instances $d$, list of sequences $S$ 
        \State $\mathit{sorted\_sequences} \gets$ Sort $S$ in descending order by length, 
        \State Initialize $\mathit{new\_batches}$ as a priority queue that sort the batches based on the sum of sequence lengths, 
        \For{$i=1 \rightarrow d$}
        \State $B_i \gets \emptyset$, $\mathit{new\_batches.\text{push}(B_i)}$
        \EndFor
        \For{$s \in sorted\_sequences$}
        \State $\mathit{new\_batches.\text{top}().\text{push}(s)}$
        \EndFor
        \State \Return $\mathit{new\_batches.\text{tolist}()}$
    \end{algorithmic}
    \end{footnotesize}

\end{algorithm}

In the case of no padding, we adopt the improved greedy algorithm to solve this problem, which is a 4/3-approximation algorithm, as shown in Algorithm~\ref{alg:rmpad}. For the $L_i$ with paddings, we propose an approximation algorithm ~\ref{alg:pad}, which combines binary and greedy approaches. The computational complexities of these two algorithms are, respectively, $O(n \log n)$ and $O(n \log (nC))$, where $C$ is the range of binary searching.

\begin{algorithm}[b!]
\caption{Post-Balancing Algorithm with Paddings}\label{alg:pad}
    \begin{footnotesize}
    \begin{algorithmic}[1]
    
        \Require count of DP instances $d$, list of sequences $S$
        \State $\mathit{sorted\_sequences} \gets $ Sort $S$ in ascending order by length,
        \Function {GetLeastBatches}{$b$} \textcolor{ACMGreen}{\# Upper bound $b$ for batch lengths}
        \State $\mathit{ret\_batches} \gets \{\{\}\}$
        \For{$s \in sorted\_sequences$}
        \If{$(\text{len}(\mathit{ret\_batches}[-1]) + 1) * s.\text{length} > b$}
        \State $\mathit{ret\_batches}.\text{push}(\{\})$
        \EndIf
        \State $\mathit{ret\_batches}[-1].\text{push}(s)$
        \EndFor
        \State \Return $\mathit{ret\_batches}$
        \EndFunction
        \State $left \gets sorted\_sequences[0].\text{length}$, 
        \State $right \gets sorted\_sequences[0].\text{length} * (\frac{n}{d} + 1)$
        \While{$left < right$} \textcolor{ACMGreen}{~\# Binary search}
            \State $mid \gets \left\lfloor \frac{left + right}{2} \right\rfloor$
            \State $\mathit{new\_batches} \gets $ \Call{GetLeastBatches}{$mid$}
            \State $right \gets \text{if } \text{len}(\mathit{new\_batches}) \leq d \text{ else } left \gets mid + 1$
        \EndWhile
        
        \State \Return $\mathit{new\_batches}$

    \end{algorithmic}
    \end{footnotesize}
\end{algorithm}

Besides, several Post-Balancing algorithms for other model architectures and the scenario where the assumption $\beta \ll \alpha$ is not valid are presented in Appendix~\ref{app:other_alg}. To sum up, our Post-Balancing algorithm completely takes effect after DP instances have randomly sampled the mini-batches. Therefore, the Post-Balancing algorithm won't violate the principle of batch randomness at all, compared with previous Pre-Balancing methods. Meanwhile, due to the ability of the Post-Balancing algorithm to perform load balancing over a wider range (across the mini-batches of all DP instances), it not only achieves a better balancing effect, but also has additional benefits, such as reducing the redundant paddings.

\subsection{Node-wise All-to-All Communicator}
\label{sec:ata}
In most current DP frameworks, each DP instance independently sample a mini-batch from a split of the training dataset, which necessitates communication across DP instances. Consequently, we devise the implementation of the Post-Balancing dispatcher, with an efficient communicator referred to as \textbf{Node-wise All-to-All Communicator}, to reduce the communication overhead and memory occupancy of examples' rearrangement. 

\subsubsection{All-to-All Batch Communicator}

\noindent \textbf{Strawman solution.} A trivial approach is to carry out an All-Gather operation on each DP instance to collect all the mini-batches and then execute the Post-Balancing algorithm. 
However, this approach will bring a substantial increase in communication overhead within the training system. Due to the fact that the communication volume of a mini-batch is proportional to its batch length $L_i$, communication overhead of this collective communication operation can be given as:
\begin{equation}
\label{equ:ag}
O_\text{All-Gather} \propto (d-1) \max_{0 \leq i < d} (L_i)
\end{equation}
where the proportional relationship is deduced by the ring-based algorithm. The communication overhead of this approach increases proportionally with the scale-up of a distributed training system, which impedes the scalability of such an approach. Moreover, mini-batches gathered from all DP instances will occupy a considerable amount of memory in the physical memory of each DP instance, which is unacceptable in the context of large-scale distributed training.

\begin{figure}[b]
  \centering
  \vspace{-0.1in}
  \includegraphics[width=\linewidth]{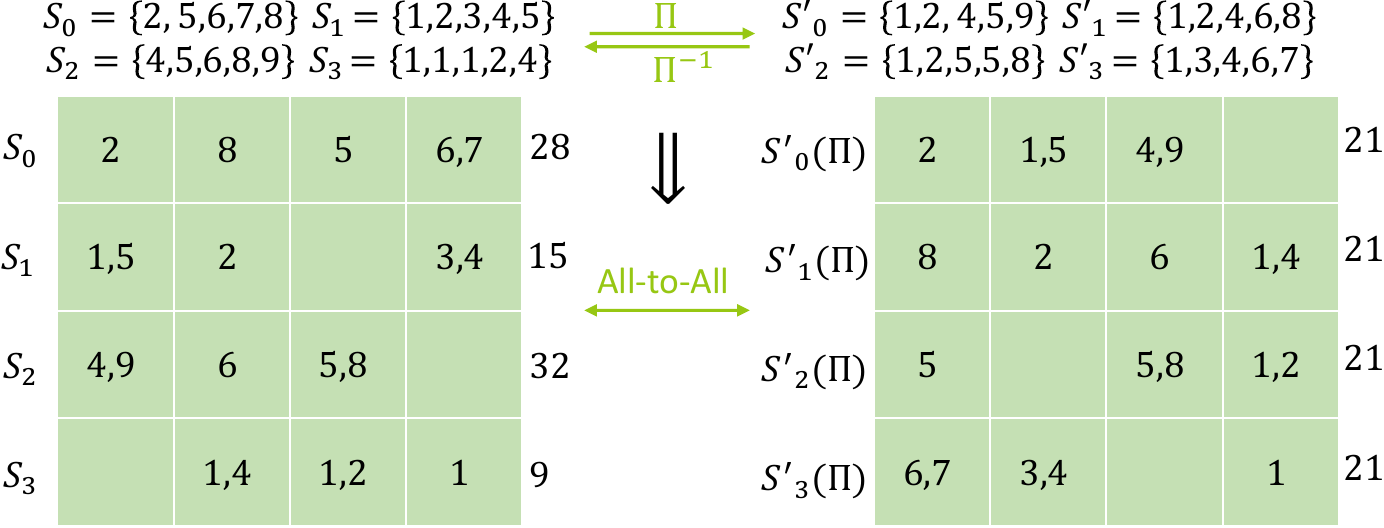}
  \vspace{-0.15in}
  \caption{An illustrative diagram exhibiting the rearrangement process using the All-to-All operation, where each number refers to the length of a sequence in batches $S_i$.}
  \label{fig:ata}
\end{figure}

According to Section~\ref{sec:alg}, the only factor that influences the solution to an Post-Balancing algorithm is the distribution of sequence lengths within all mini-batches, i.e. $l_{i,j}~ (0 \leq i < d, 0 \leq j < b_i)$. Thus, it's sufficient to only communicate all the $l_{i,j}$ across DP instances with the All-Gather operation, which incurs almost negligible communication overhead. Then, we could execute the Post-Balancing algorithm on each instance and obtain the optimal rearrangement $\Pi$, which mappings each example from the source instance to the destination instance. This rearrangement can be implemented by the collective communication operation, All-to-All, as shown in Figure~\ref{fig:ata}. Owing to the point-to-point communication protocol of All-to-All, the communication overhead of this approach can be denoted as (Detailed deduction of Equation~\ref{equ:ag}, ~\ref{equ:ata} and~\ref{equ:node} can be found in Appendix~\ref{app:deduct}):
\begin{equation}
\label{equ:ata}
O_\text{All-to-All} \leq \Omega_\text{All-to-All} \propto \max_{0 \leq i < d} (L_i) 
\end{equation}
where $\Omega_\text{All-to-All}$ refers to the upper bound of communication overhead of All-to-All and is deduced based on the point-to-point communication protocol.
From the equation, it can be observed that the communication overhead of All-to-All does not increase with the scale-up of the cluster anymore. Moreover, this approach barely requires instances to allocate redundant memory in the memory, which is much more favorable to memory utilization.

\subsubsection{Node-wise Rearrangement Algorithm}
\label{sec:nw}

In the large-scale cluster, there exists heterogeneity in communication topologies between intra-node and inter-node instances, as shown in Figure~\ref{fig:top}, and the proportionality factor in~\eqref{equ:ata} is determined by the minimum point-to-point communication bandwidth. Due to the significant disparity between inter-node and intra-node point-to-point bandwidths (e.g. intra-node communication using NVlink typically offers hundreds of GBs point-to-point bandwidth, whereas inter-node communication via Ethernet usually allocates merely dozens of GBs bandwidth per instance), the communication overhead of All-to-All communicator is determined by the maximum volume of inter-node communication:

\begin{figure}[b]
  \centering
     \vspace{-0.1in}
  \includegraphics[width=\linewidth]{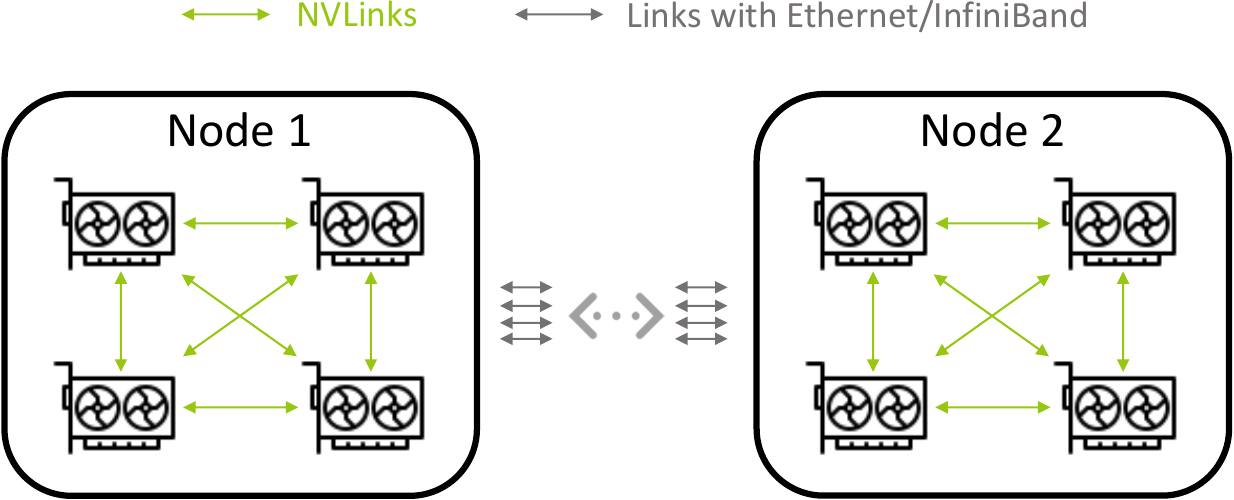} 
  \vspace{-0.2in}
  \caption{The heterogeneous communication topology in the large-scale cluster for distributed training. }
  \label{fig:top}

\end{figure}

\begin{equation}
\label{equ:node}
O_\text{All-to-All} \propto \max_{0 \leq i < d} (\underset{i'(\Pi) \notin N(i)}{\sum} l_{i,j}) 
\end{equation}
where $i'(\Pi)$ is the new instance of $l_{i,j}$ under the rearrangement $\Pi$ and $N(i)$ refers to the set of instances on the same node with the $i$-th instance. We can find that, during the All-to-All operation, pairs of instances from different nodes become stragglers compared with intra-node pairs.

\begin{algorithm}[t!]
\caption{Node-wise Rearrangement Algorithm}\label{alg:nwa}
    \begin{footnotesize}
    \begin{algorithmic}[1]
        \Require count of DP instances $d$, count of DP instances on a node $c$, \
        original rearranged batches $\{S'_0, \ldots, S'_{d-1}\}$
        \State $\mathit{cost\_matrix} \gets \{\{0\} \times d\} \times d$, 
        \For{$i = 0 \rightarrow d-1$}
            \For{$l \in S'_i$}
                \State $\mathit{cost\_matrix}[l.\text{from}][i] += l.\text{length}$
            \EndFor
        \EndFor

        \State $\mathit{x} \gets \text{Variable}(\frac{d}{c}, c)$, $max\_cost \gets \text{Variable(1)}$
        \State $cons \gets [\text{sum of each column of } x \text{ is } c, \text{sum of each row of } x \text{ is 1}]$
        \For{$i = 0 \rightarrow \frac{d}{c} - 1$}
            \State $k \gets i \times c$
            \State $cons.append(\mathit{cost\_matrix}[k : k + c] \cdot (1 - x[:, i]) \leq max\_cost)$
        \EndFor
        
        \State $prob \gets \text{Problem}(\text{Minimize}(max\_cost), cons)$
        \State $prob.solve()$

        \State $permuted\_batches \gets $ Permute $\{S'_0, \ldots, S'_{d-1}\}$ with $x$.value
        \State \Return $\mathit{permuted\_batches}$
    \end{algorithmic}
    \end{footnotesize}
\end{algorithm}

Therefore, we propose \textbf{Node-wise Rearrangement Algorithm} to further reduce the communication overhead.
For a given set of mini-batches $S_i (0 \leq i < d)$, the rearrangement $\Pi$ solved by any Post-Balancing algorithm can be instantiated as an ordered set, $\Pi \equiv (S'_0(\Pi), ..., S'_{d-1}(\Pi))$, as shown in Figure~\ref{fig:ata}. For the given $\Pi$, we can calculate the matrix $V$, where the $(i, j)$-entry represents the communication volume between $i$-th and $j$-th DP instances. Apparently, permutation on the ordered set leads to the same permutation on columns of $V$, which further induces changes of the communication overhead. Meanwhile, due to the fact that the objective in Post-Balancing algorithms is regardless of the order of $\Pi$, any permutation on the ordered set is invariant for the objective. Hence, we formulate the objective of Node-wise Rearrangement Algorithm as:
\begin{equation*}
 \text{Objective:} \quad  \underset{P}{\text{minimize}} ~ \max_{0 \leq i < d} (\underset{i'(\Pi') \notin N(i)}{\sum} l_{i,j})
 \end{equation*}
 where $P$ refers to the permutation on the rearrangement $\Pi$ and new rearrangement $\Pi' = P(\Pi)$. This objective implies that it's feasible to reduce the communication overhead of All-to-All by assigning more communication volume to intra-node communication instead of inter-node communication.

As shown in Algorithm~\ref{alg:nwa}, we leverage Integer Linear Programming (ILP) to devise Node-wise Rearrangement Algorithm, which involves $O(d^2/c)$ variables ($c$ is the count of DP instances on a node) and $O(d)$ constraints, thereby introducing overhead of  tens of milliseconds on a large-scale cluster. In the workflow of MLLM training, however, the overhead can be overlapped, as detailed in following Section~\ref{sec:COO}. Additionally, Node-wise Rearrangement Algorithm is applicable to all the Post-Balancing algorithms, because it operates solely on the solutions provided by these algorithms, which obviates the necessity for bespoke modifications tailored to the implementation of particular Post-Balancing algorithm.

\section{MLLM Global Orchestrator}
\label{sec:glo}

\begin{figure}[t]
  \centering
  \includegraphics[width=\linewidth]{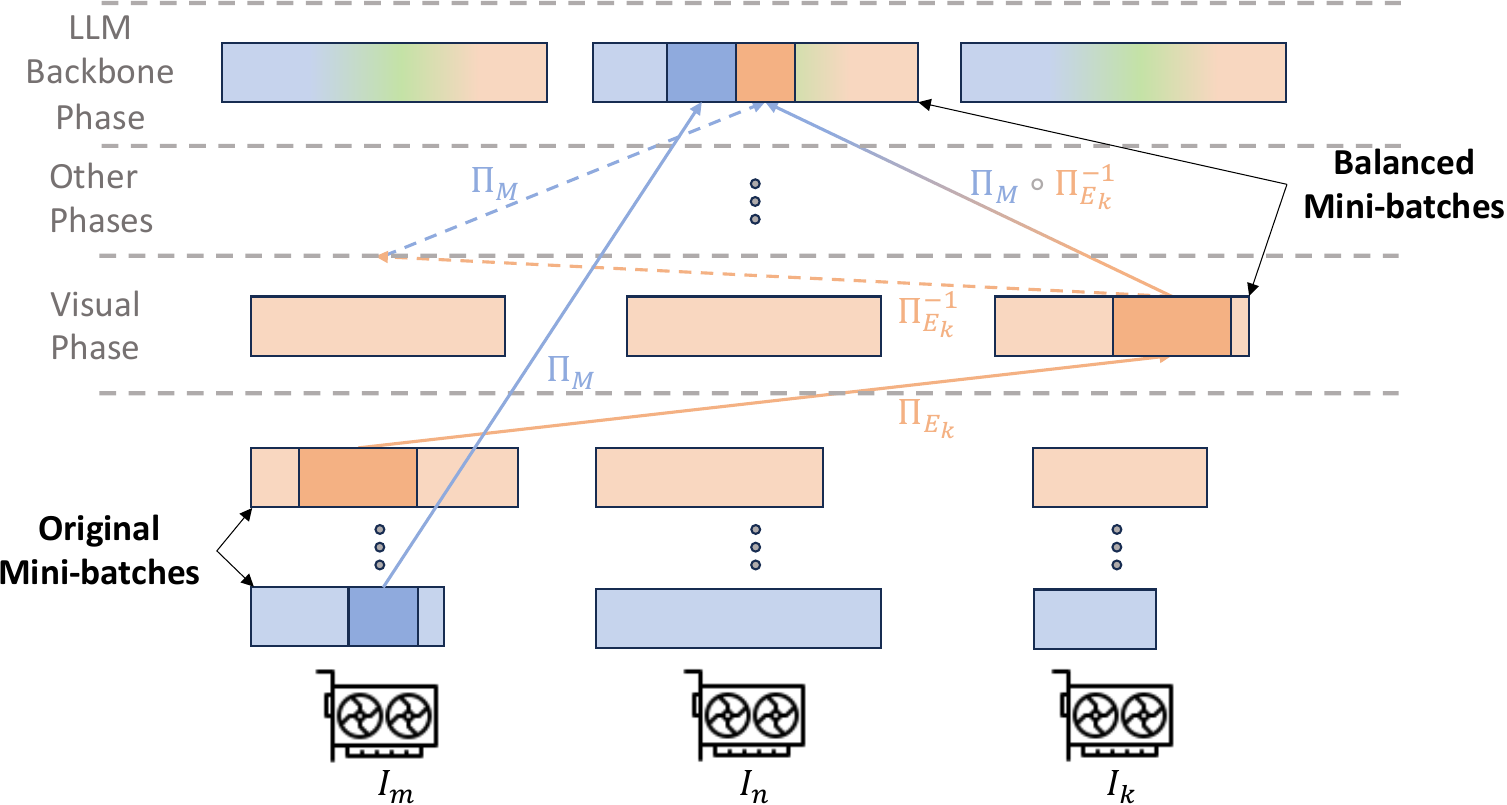}
 \vspace{-0.2in}
  \caption{The diagram that illustrates the data flow of an example in MLLM Global Orchestrator. }
  \label{fig:global}
  
 \vspace{-0.15in}
\end{figure}

After the preparation on the training of single-modality data, we can refocus on the workflow of MLLM training. In order to eliminate the imbalance in mini-batches in each phase, a feasible approach is to carry out the Batch Post-Balancing for each phase, on the corresponding data to be processed. However, it's neither reasonable nor efficient to straightforward apply the Post-Balancing Dispatchers separately, due to the data dependencies in MLLM training. We devise the \textbf{MLLM Global Orchestrator} for the workflow of MLLM training to guarantee the correctness and efficiency of Batch Post-Balancing during training, and finally make an efficient framework of MLLM training realizable.

\medskip\noindent\textbf{Subsequences assembly.} In the workflow of MLLM training, the phase of LLM backbone is distinct, compared to other phases of encoders, due to additional data dependencies. For a given example in the training, the encoded results from different encoders, as subsequences, are assembled and interleaved into a sequence according to a predefined order. Therefore, to perform the Post-Balancing algorithm in MLLM Global Orchestrator, we set the sequence length of an example $l_{i,j}$ as the length of the whole interleaved sequence, instead of the lengths of texts in this example.

 
Moreover, the obtained rearrangement, $\Pi_M$, maps the examples from their original instances to the destination instances where the LLM backbone will process their interleaved sequences. Therefore, the rearrangement of texts is straightforward, because texts are just located on the original instances. As for the encoded results which derive from the metadata of corresponding modalities, because the mini-batches of metadata are formed after the rearrangement of encoder dispatchers, most of the encoded results aren't located on their original instances. Consequently, straightforward applying the rearrangement $\Pi_M$ to the encoded results fails to rearrange the subsequences of examples to their destinations. There, A trivial approach is to reset the encoded results to their original instances with the inverse rearrangement $\Pi^{-1}$ and then apply the rearrangment $\Pi_M$ to assemble the subsequences on the destination instance.

\medskip\noindent\textbf{Rearrangement composition.}  Though resetting the encoded results to their original instances is able to guarantee the correctness during training, it still leads to higher communication overhead due to more All-to-All operations. 
We denote the rearrangement obtained from the Post-Balancing dispatchers of encoders as $\Pi_{E_k}$, where $E_k$ represents the $k$-th encoder of MLLM. If we reset the encoded results, $A_{E_k}$, of the $k$-th encoder to their original instances and then apply the rearrangement $\Pi_L$, we will get the rearranged encoded results, which will serve as the subsequences on the corresponding instances, denoted as $A'_{E_k}$: 
$$
A'_{E_k} = \Pi_M\left(\Pi_{E_k}^{-1}(A_{E_k})\right).
$$

Due to the fact that these two linear mappings, $\Pi_M$ and $\Pi_{E_k}^{-1}$, satisfy the associative law, we can first compound them into a single linear mapping, $\Pi_M \circ \Pi_{E_k}^{-1}$, and apply it to the encoded results $A_{E_k}$ simultaneously, i.e.
$$
A'_{E_k} = \left(\Pi_M \circ \Pi_{E_k}^{-1} \right)(A_{E_k}).
$$
By {Rearrangement Composition}, we can integrate two All-to-All operations into a single one for each encoder in the forward pass. Moreover, because each rearrangement between the encoders and the LLM backbone will be accompanied by a rearrangement in the backward pass, we can reduce the communication overhead by half in total, which accelerates the speed of MLLM training.

\medskip\noindent\textbf{Computation overhead overlapping.}
\label{sec:COO}
For arbitrary dispatchers, we can split the execution of a dispatcher into two parts, respectively, referred to as computation and communication. The former is primarily encompassed by the execution of Post-Balancing algorithm and Node-wise Rearrange algorithm, which are both executed by the CPUs (also includes relatively lightweight operations, like Rearrangement Composition in MLLM Global Orchestrator). The latter carries out the All-to-All operation to rearrange the mini-batches practically across DP instances. 

To avoid interference between communication of dispatchers with original communication of distributed training, we tend to insert the communication into the forward pass, which is the critical path of training, instead of paralleling it with the forward pass. In contrast, the computation can overlap with the forward pass. Because inputs to the algorithms of computation are sequence lengths of all examples in mini-batches, which are predictable owing to the characteristic of MLLM, the computation can be executed as soon as all the mini-batches have been sampled. Moreover, the sampling of mini-batches runs in parallel with the forward pass through prefetching. To allow the computation of dispatchers to parallel with the forward pass, we can integrate it into the prefetching process, which also guarantees that the terminal rearrangements are solved before carrying out the All-to-All operations. In this way, we can completely overlap the computation overhead of Post-Balancing Dispatchers, especially these algorithms that are computing-intensive, further reducing the overall overhead.


\section{Implementation}
We implement a system incorporating OrchMLLM, leveraging the feature provided by PyTorch 2.0, Fully Sharded Data Parallel (FSDP), which is a well-established framework with fine efficiency and scalability. Besides, the system inherits the universality and applicability of FSDP, and allows convenient adaptation to various modalities and models. We implement the entire system from scratch which comprises 5.1k lines of codes in Python and C++. 

\medskip\noindent \textbf{Batch Post-Balancing Algorithms.}
All Batch Post-Balancing Algorithms are integrated into the Batch Post-Balancing Dispatcher and will be selected according to the specified balance policy. To reduce the overhead of executing the balance algorithms, instead of Python, we implement all algorithms, presented in Section~\ref{sec:alg} and ~\ref{app:other_alg}, with C++ and link them to the Python code with pybind.

\medskip\noindent \textbf{Node-wise All-to-All Communicator.}
Node-wise All-to-All communicator is implemented based on PyTorch Distributed~\cite{torchdist} library with NCCL as the communication backend. 
The Node-wise Rearrange Algorithm is implemented with python package {CVXPY}, using the solver {CBC}.

\medskip\noindent \textbf{MLLM Global Orchestrator.}
 We define a structure to record some details, including the counts of subsequences of different modalities and the order in which the subsequences are interleaved, for each example, which are gathered for MLLM Global Orchestrator to carry out Batch Post-Balancing. Moreover, we refine the dataloader in OrchMLLM to integrate the computation part of dispatchers into prefeching.
\section{Evaluation}
In this section, we first use large-scale experiments to demonstrate the overall performance improvements of OrchMLLM over Megatron-LM and the baseline of us. Subsequently, I conducted ablation and comparative experiments on several components within OrchMLLM, based on microbenchmarks.

\medskip\noindent \textbf{Setup.} Our experiments are conducted on a production GPU cluster for MLLM training, with each node equipped with eight NVIDIA H100 GPUs, 1.8TB of memory, and 88 vCPUs. GPUs within one node are interconnected by 900GB/s (bidirectional bandwith) NVLink, while nodes are connected by 8*400 Gbps RDMA network based on InfiniBand. The experiments for overall results use the same cluster with 2560 GPUs, and the microbenchmark utilizes 128 GPUs. We use PyTorch 2.4.0 and NVIDIA CUDA 12.4 to build the system and for our evaluation.

\begin{table}[t]
    \centering
    \caption{
        The configurations of submodules.
    }
     \vspace{-0.1in}
    \label{tab:sizes}
    \footnotesize
    \begin{tabular}{c|ccccc}
    \toprule
     &  &  &  &\textbf{FFN} & \textbf{Total} \\
    \textbf{Models} & \textbf{Sub-} & \textbf{\# of} & \textbf{Hidden} &\textbf{Hidden} & \textbf{Para-} \\
    & \textbf{modules} & \textbf{Layers} & \textbf{Size} &\textbf{Size} &\textbf{meters} \\
    \midrule
                                & LLM    &  $28$  & $3584$ & $18944$ & $7$B \\
    MLLM-10B                    & Vision &  $36$  & $2048$ & $8192$  & $2$B \\
                                & Audio  &  $32$  & $1280$ & $5120$  & $0.6$B \\
    \midrule
                                & LLM    &  $48$  & $5120$ & $13824$ & $14$B \\
    MLLM-18B                    & Vision &  $40$  & $2400$ & $9600$  & $3$B  \\
                                & Audio  &  $32$  & $1280$ & $5120$  & $0.6$B \\
    \midrule
                                & LLM    &  $80$  & $8192$ & $29568$ & $72$B \\
    MLLM-84B                    & Vision &  $45$  & $3200$ & $12800$ & $6$B  \\
                                & Audio  &  $48$  & $3072$ & $12288$ & $6$B  \\
    \bottomrule
    \end{tabular}
     \vspace{-0.1in}
\end{table}

\medskip\noindent \textbf{Models.} For the LLM backbone, we choose the architecture of Qwen2\cite{yang2024qwen2technicalreport}, which also serves as the backbone both in Qwen2-VL~\cite{wang2024qwen2vlenhancingvisionlanguagemodels} and Qwen2-Audio~\cite{chu2024qwen2audiotechnicalreport}. Targeting visual and auditory modalities, we respectively adopt ViT~\cite{dosovitskiy2020image} for the visual encoder and the encoder of Whisper~\cite{whisper} for the auditory encoder, two widely established models for image understanding and audio comprehension. We choose varying configurations of encoders to match differently sized LLM backbones, as shown in Table~\ref{tab:sizes}.
The three types are designated by the total parameter count of submodules, respectively: MLLM-10B, MLLM-18B, and MLLM-84B. The connectors between the encoders and the LLM backbone are universally MLPs. In additional, a downsample operation for the encoded results will be carried out before the connectors, and the downsample rates are respectively set as 1, 4, 4 for the visual results and 2, 2, 4 for the auditory results.


\medskip\noindent \textbf{Datasets}. As discussed in Section~\ref{sec:mod}, we integrate several open-sourced datasets into the dataset for evaluation. For the visual modality, we adopt the instruction tuning dataset of LLaVA-1.5\cite{Liu_2024_CVPR}, which encompasses varieties of visual tasks. For the auditory modality, we combine Librispeech\cite{7178964}, a dataset for Automatic Speech Recognition (ASR), with AIR-Bench which is an integrated dataset for Speech Question Answering. We generate training data by randomly sampling the example from the whole dataset, ensuring their random distribution across two types of datasets. As profiled and analyzed in Section~\ref{sec:mod}, Modality Composition Incoherence emerges in this dataset.

\medskip\noindent \textbf{Input preprocessing.} The upper bounds for image resolutions of the three MLLMs are set at 448, 672, and 896, all with the same patch size 14. Only images that are larger than the upper bound will be resized, and the sizes of preprocessed images are dynamic. The patches of the images are batched along the sequence length, with no padding in the batch. The audio sample rate is fixed at 16000 and the sequences of audios are batched with paddings, due to the existence of the convolution architecture in the auditory encoder. The sequences for the LLM backbone are batched with no padding.

\medskip\noindent \textbf{Metrics}. We use the Model FLOPs Utilization (MFU) as the primary metric to evaluate OrchMLLM. MFU measures the percentage of GPU FLOPs that are effectively utilized during model training. Given that there are redundant computation caused by paddings, we universally calculate effective GPU FLOPs without paddings. Besides, we leverage the training throughput (TPT) to evaluate the training speed, defined as the tokens processed by the LLM backbone per second on each GPU. The GPU memory usage, defined as the maximum of the memory usage during the training process, also serves as the auxiliary metric, in the ablation studies, to demonstrate OrchMLLM's effect in optimizing memory occupancy.

\subsection{Overall Results}

\noindent \textbf{OrchMLLM setup.} Implemented based on FSDP, we adopt the whole cluster (2560 H100s) as the data parallel group and set the hybrid group size\cite{zhao2023pytorch} for ZeRO3 as 256. We set the mini-batch size according to the memory usage during training, to avoid the error of OOM, respectively 80, 60, 30 for three MLLMs.

\begin{figure}[b]

     \vspace{-0.1in}
  \centering
  \includegraphics[width=\linewidth]{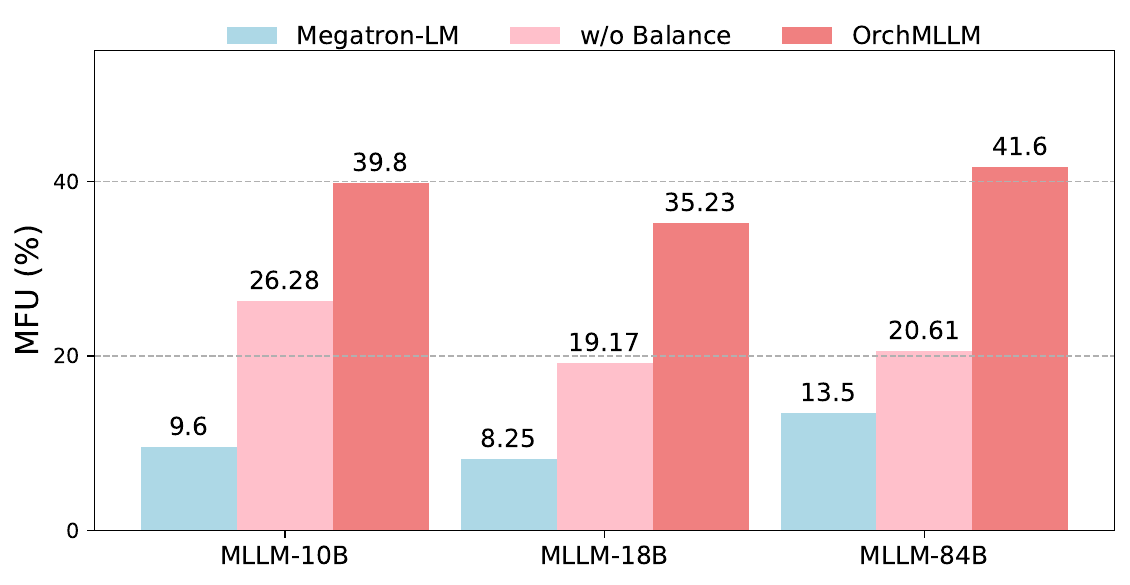}
  \caption{The overall results of MFU.}
    \label{fig:mfu}
\end{figure}
\medskip\noindent\textbf{Baseline setup.} We adopt two baselines in this part to demonstrate the efficiency and scalability of our method. The first is Megatron-LM, an established training framework for LLM. We retrofit the workflow of training text-image MLLM in the Megatron-LM to support the training of MLLM shown in Table~\ref{tab:sizes}, by integrating the auditory encoder into the framework and enabling the pipeline parallelism of MLLM with two or more encoders (without any other retrofits). The PP sizes for three MLLMs are 2, 4, 10, while the TP size is universally 8.
The global batch sizes are respectively set as 5120, 5120, 2560. The second baseline is the OrchMLLM without any balancing, which is used to isolate the impact of system implementation and exemplify the effectiveness of our method.
 We set the mini-batch size according to the same standard, respectively: 65, 40, 15 for three MLLMs and other setups remain consistent.
 
 It should be noted that we omit direct comparison with the contemporary framework, DistTrain, as its implementation is closed-sourced. Meanwhile, for the scope of this paper (i.e., handling mini-batches imbalance in MLLM training), DistTrain, according to the analysis in \S~\ref{sec:existing}, is effectively a variant of Pre-balancing methods, with which we compare in \textbf{Comparison with Pre-Balancing methods} (\S\ref{sec:micro}). 

 The experimental results are shown in Figure~\ref{fig:mfu} and Figure~\ref{fig:tpt}, respectively for MFU and the training throughput. From the experimental results, we can draw the following conclusions:

\begin{figure}[t]
  \centering
  \includegraphics[width=\linewidth]{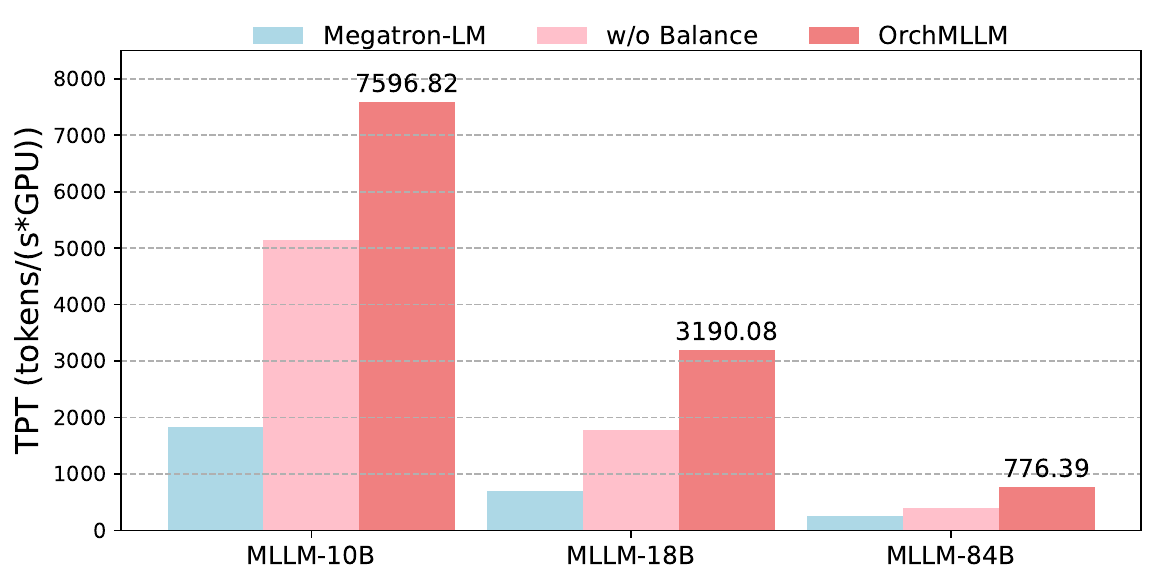}
  \caption{The overall results of the training throughput in tokens.}
    \label{fig:tpt}
   \vspace{-0.2in}
\end{figure}

\renewcommand{\labelitemi}{\textbullet}
 \begin{itemize}[leftmargin=10pt]
\item From Figure~\ref{fig:mfu}, OrchMLLM achieves $41.6\%$ MFU on the large-scale cluster with 2560 H100s. Due to the performance disparity between the H100 and A100, this result is roughly equivalent to $60\%$ MFU on the same scaled cluster of A100s, and approaches the state-of-the-art efficiency of LLM training, which is the theoretical upper limit for MLLM training. The results on such a large-scale cluster demonstrate high efficiency and scalability of OrchMLLM.

\item Compared with Megatron-LM, OrchMLLM achieves significant breakthroughs. OrchMLLM outperforms Megatron-LM with $3.1-4.1\times$ the MFU and $3.1-4.2\times$ the training throughput, though the performance of Megatron-LM also suffers from the model heterogeneity~\cite{zhang2024disttrainaddressingmodeldata}. In additional, these breakthroughs highlight the potential and feasibility of leveraging FSDP for training MLLMs to avoid model heterogeneity.

\item The contrastive experiment prominently demonstrate the effectiveness of our method. OrchMLLM outperforms OrchMLLM without balance with $1.5-2.0\times$ the MFU and $1.4-1.9\times$ the training throughput. The ratios between them increase as MLLM grows larger, because larger models exert greater pressure on the GPU memory and advantages in memory utilization that OrchMLLM gains through balancing become more pronounced, further underscoring the efficacy and necessity of the method.

\end{itemize}

\subsection{Overhead Analysis}

The Post-Balancing method is on the critical path during the forward pass, so it is needed to analyze the overhead in the system. We adopt MLLM-10B, set the mini-batch size as 60, and profile the overhead in latency, including communication overhead and some extra durations, on differently sized clusters from 64 to 2560. Meanwhile, we also profile the average duration of the forward pass to estimate the overhead's proportion in the training process.
\begin{table}[t]
    \centering
    \caption{
        The profiling results of OrchMLLM.
    }
     \vspace{-0.1in}
    \label{tab:overhead}
    \footnotesize
    \begin{tabular}{c|cccccc}
    \toprule
     GPUs & 64 & 128 & 256 & 512 & 1024 &2560 \\
     \midrule
    Overhead (ms)               & $16.66$ &  $18.49$  & $21.32$ & $24.64$  & $31.37$ & $53.88$ \\
    \midrule
    Duration (s)                    & $3.79$ &  $3.81$  & $3.85$ & $3.92$  & $3.96$ & $4.05$  \\
    
    \bottomrule
    \end{tabular}
     \vspace{-0.15in}
\end{table}

From the Table~\ref{tab:overhead}, it is evident that the overhead of OrchMLLM consistently remains dozens of milliseconds, accounting for less than $2\%$ in the duration of the forward pass. Meanwhile, in the backward pass, the overhead introduced by OrchMLLM is the communication overhead in the backward process of All-to-All operations, which is lower than that in the forward pass (due to more rounds of communication). Therefore, we can conclude that the overhead introduced by OrchMLLM constitutes only a negligible fraction of the training process, which further exemplifies the scalability of OrchMLLM.

\subsection{Ablations and Microbenchmarks}
\label{sec:micro}
The ablation studies and microbenchmarks in this part are all performed on the cluster with 128 H100s. The mini-batch sizes are respectively set as: 75, 50, 25 for three MLLMs without further specification. Due to the strict proportional relationship between MFU and training throughput, we don't leverage training throughput as metrics anymore.

\medskip\noindent \textbf{Comparison with Pre-Balancing methods.}
As we have analyzed in Section~\ref{sec:existing}, existing Pre-Balancing methods are only capable of addressing imbalances in a single modality and a common approach is to guarantee balance only during the phase of LLM backbone. Therefore, we ablate the balancing dispatchers for the phases of encoders and further exemplify the superiority of OrchMLLM over existing Pre-Balancing methods with the ablation experiments.

\begin{figure}[b]
    \centering
         \vspace{-0.1in}
    \begin{subfigure}{0.48\linewidth} 
        \centering
        \includegraphics[width=\linewidth]{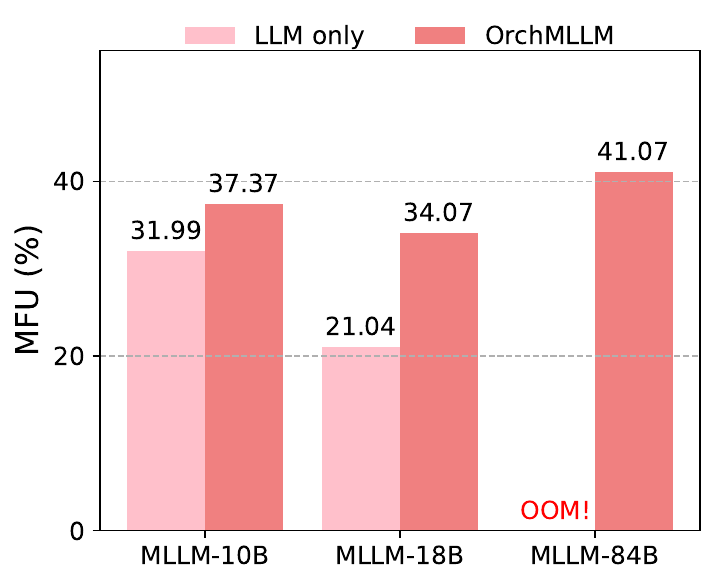} 
        \caption{MFU}
        \label{fig:subfig1}
    \end{subfigure}
    \hfill 
    \begin{subfigure}{0.48\linewidth} 
        \centering
        \includegraphics[width=\linewidth]{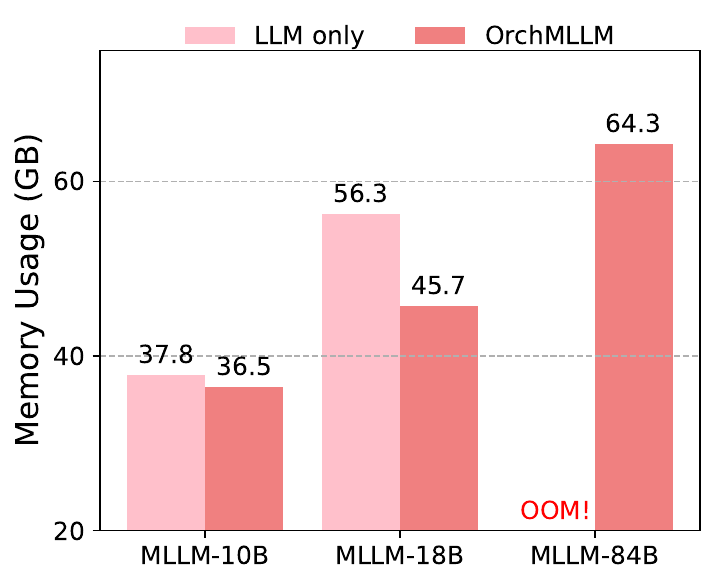} 
        \caption{GPU Memory Usage}
        \label{fig:subfig2}
    \end{subfigure}-
     \vspace{-0.1in}
    \caption{The ablation results of encoder balancing and comparison with Pre-Balancing methods.}

    \label{fig:llm}
\end{figure}

From Figure~\ref{fig:llm}, we can find that OrchMLLM consistently outperforms OrchMLLM with only LLM balance in terms of both MFU and GPU memory usage. Moreover, the advantages become increasingly pronounced as the size of MLLM grows, to the extent that OrchMLLM with only LLM balance even triggers an OOM error for the MLLM-84B. (As a reference, it can run smoothly with the mini-batch size set to 18, achieving an MFU of $24.16\%$ and a GPU memory usage of $62.7$GB.) This indicates that Modality Composition Incoherence indeed causes imbalance even in the case of LLM balance and imbalance in the phases of encoders has a more significant impact on GPU utilization as model size increases. From this ablation experiments, it can be concluded that, compared with these Pre-Balancing methods, the introduction of Post-Balancing algorithms along with OrchMLLM is necessary for comprehensively eliminating the imbalance in MLLM training and unleashing the potential of accelerators, especially for the MLLMs with larger capacity.

\medskip\noindent \textbf{Post-Balancing algorithms}
In Section~\ref{sec:alg}, we tailor several Post-Balancing algorithms for different scenarios, including the algorithms for different batching strategies, i.e. with paddings or not. In OrchMLLM, we adopt different balancing algorithms for the phases of the vision encoder and the auditory encoder according to the batching strategies. In this part, we conducted two sets of contrastive experiments by changing the balancing algorithm of one phase to that of the other, respectively referred to as \textit{all rmpad} and \textit{all pad} as follows. The algorithm for the phase of the LLM backbone remains consistent. 

\begin{figure}[t]

  \centering
  \includegraphics[width=\linewidth]{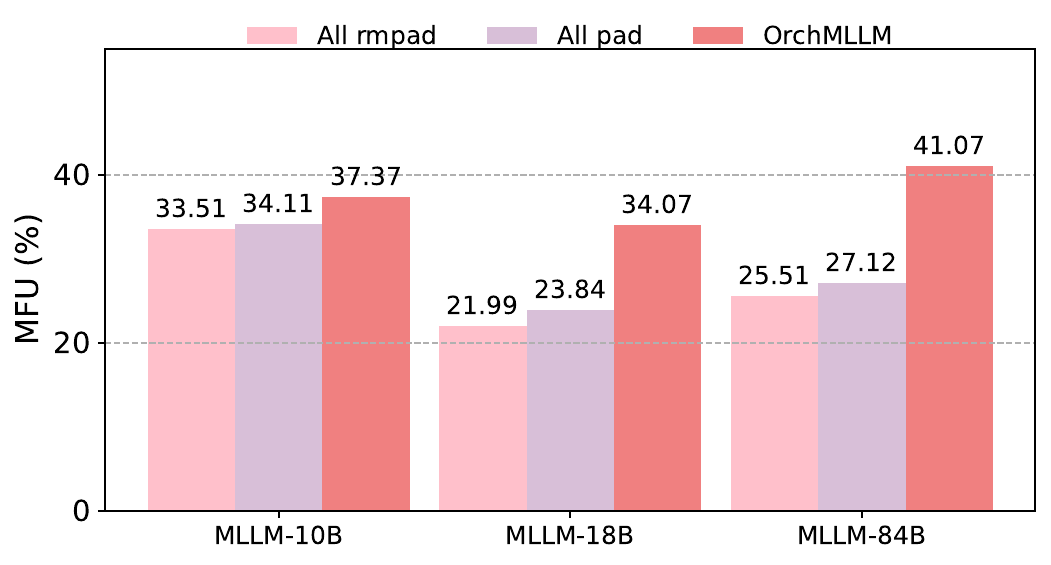}
  \caption{The results of two experiments with rigid algorithms, compared with OrchMLLM with tailored algorithms.}
  \label{fig:policy}
   \vspace{-0.15in}
\end{figure}

As shown in Figure~\ref{fig:policy}, the MFUs of these two groups are significantly lower than that of OrchMLLM. This indicates that a single algorithm for all phases of encoders is not effective enough to eliminate the imbalance in MLLM training due to the complicate scenarios derived from divergent model architectures of various encoders and diverse distributions of sequence lengths. This microbenchmark demonstrates the necessity of tailoring several Post-Balancing algorithms.

\begin{figure}[b]
 \vspace{-0.15in}
    \centering
    \begin{subfigure}{0.48\linewidth} 
        \centering
        \includegraphics[width=\linewidth]{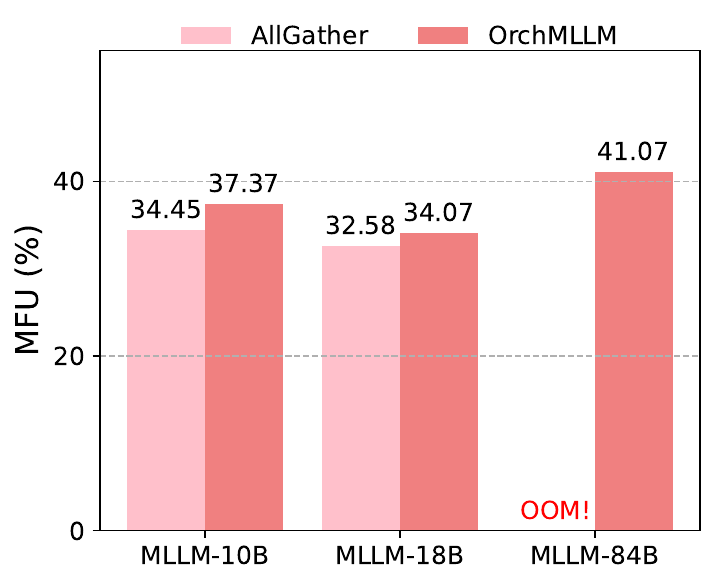} 
        \caption{MFU}
        \label{fig:subfig1}
    \end{subfigure}
    \hfill 
    \begin{subfigure}{0.48\linewidth} 
        \centering
        \includegraphics[width=\linewidth]{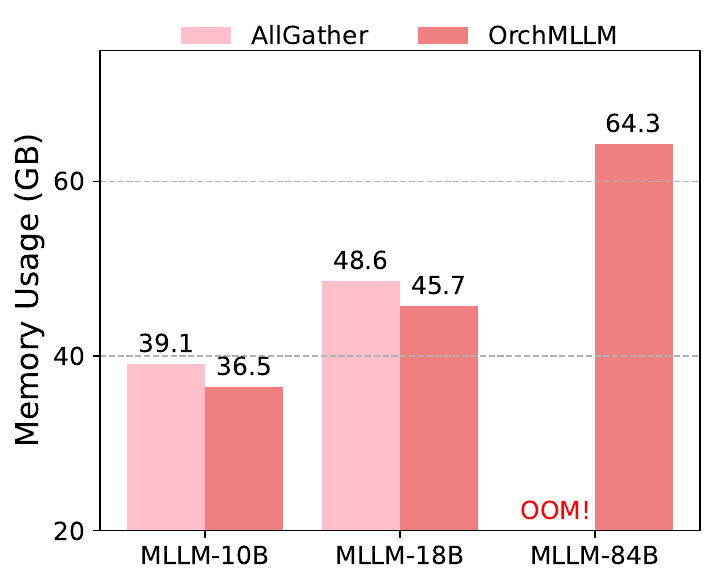} 
        \caption{GPU Usage}
        \label{fig:subfig2}
    \end{subfigure}
   \vspace{-0.1in}
    \caption{The comparative results of AllGather.}
    \label{fig:combined_fig}
\end{figure}

\medskip\noindent \textbf{Node-wise All-to-All Communicator.}
In Section~\ref{sec:ata}, we introduce Node-wise All-to-All Communicator as an efficient component of Batch Post-Balancing Dispatcher. Firstly, we substitute the communicator in OrchMLLM with a communicator implemented with All-Gather. The experimental results are displayed as Figure~\ref{fig:combined_fig}.

\begin{figure}[t]

  \centering
  \includegraphics[width=\linewidth]{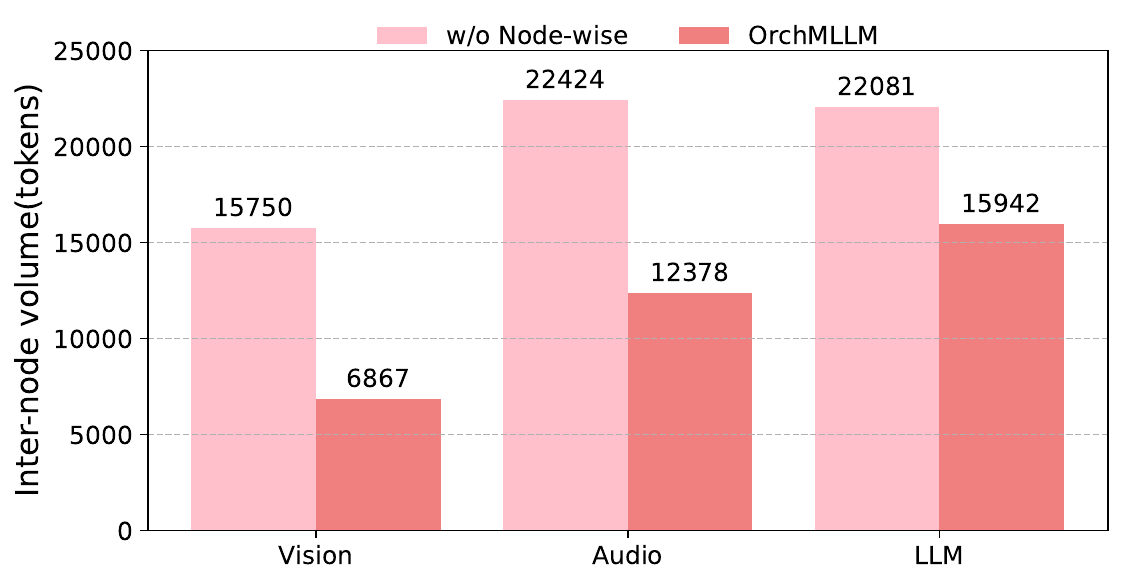}
   \vspace{-0.1in}
  \caption{The ablation results of the Node-wise Rearrange Algorithm. The metric is the average inter-node communication volume of dispatchers for each modality (per iteration).}
  \label{fig:node}
   \vspace{-0.15in}
\end{figure}

It can be concluded from the results that OrchMLLM outperforms the All-Gather communicator both in MFU and memory usage. As for the All-Gather communicator, the training of MLLM-84B crashes again because of OOM and it can accommodate a mini-batch size of 20 with an MFU of $25.51\%$ and a GPU memory usage of $61.8$GB. It is evidental that All-to-All Batch Communicator is effective in reducing both communication overhead and memory occupancy of batches' rearrangement.

Then, we ablate the Node-wise Rearrange Algorithm and compare the communication overhead with OrchMLLM. According to the analysis in Section~\ref{sec:nw}, the communication overhead is determined by the longest execution duration of All-to-All operations among DP instances, which is basically proportional to the inter-node communication volume on the DP instance. Due to the significant fluctuations in communication overhead in real-world environments, profiling becomes challenging. Therefore, we leverage the communication volume to more intuitively demonstrate the differences between two sets of experiments, as shown in Figure~\ref{fig:node}.

Under this setting, the reduction of communication overhead through Node-wise Rearrange Algorithms ranges from $0.436$ to $0.722$ for dispatchers of different modalities. Moreover, because OrchMLLM adopts tailored balancing algorithms for different phases, it is proven to be effective for different Post-Balancing algorithms, although concrete effectiveness can be influenced by the specific algorithms and data distribution. Therefore, Node-wise Rearrange Algorithm effectively reduces the overhead of dispatchers, because the communication overhead constitutes the majority of the total overhead through computation overhead overlapping.

\section{Related Work}

\noindent\textbf{MLLM paradigms.}There are broadly two ways to fuse multimodal information, namely, token-level and feature-level fusion. 
For token-level fusion, some~\cite{pandagpt, llava, pmc-vqa} simply use a MLP-based connector to bridge the modality gap, but structures with more complexity like Q-Former are also being explored~\cite{li2023blip, dai2023instructblip}.
Conversely, some works insert extra cross-attention layers (Flamingo~\cite{alayrac2022flamingo}) or expert modules (CogVLM~\cite{wang2024cogvlm}) into LLMs.
Zeng et al.~\cite{zeng2023matters} empirically reveal that the token-level fusion performs better in terms of VQA benchmarks. Both simplicity and effectiveness contribute to the popularity of token-level fusion, but the core sight of OrchMLLM can apply to both and only few refactoring on MLLM Global Orchestrator is needed.

\vspace{-0.02in}
\medskip\noindent\textbf{LLM training.}
Many efforts have been made to optimize the training of LLMs from system perspectives.
For LLM pretrain, Megatron-LM~\cite{shoeybi2019megatron} and DeepSpeed-Megatron~\cite{smith2022using} propose customized 3D-parallelism
and are de facto standards for training large LLMs.
With the proposal of sequence parallelism (SP)~\cite{li2021sequence} and expert parallelism (EP)~\cite{liu2023janus}, they are integrated into aforementioned frameworks.
DeepSpeed-ZeRO~\cite{rajbhandari2020zero} and Pytorch-FSDP~\cite{zhao2023pytorch} reduce redundant memory consumption in DP.
Fault tolerance through replication and checkpoint is advanced in large training clusters by studies~\cite{jiang2024megascale, hu2024characterization}.
Efforts like~\cite{athlur2022varuna, thorpe2023bamboo, jang2023oobleck} further optimize recovery process in cloud spot instance scenarios.
These system optimizations of LLM training are orthogonal to OrchMLLM, because OrchMLLM only operates across DP instances. 

\vspace{-0.02in}
\medskip\noindent\textbf{Multimodal model training.}
Many system optimizations have been proposed to train both small multimodal models (e.g., CLIP~\cite{radford2021learning} and LiT~\cite{zhai2022lit}) and MLLMs efficiently.
DistMM~\cite{huang2024distmm} and DistTrain~\cite{zhang2024disttrainaddressingmodeldata} tackle model heterogeneity by introducing disaggregated placement and partitioning to evenly distribute workload.
GraphPipe~\cite{jeon2024graphpipe} and Optimus~\cite{feng2024optimus} are proposed to address graph dependencies in multimodal models to minimize pipeline bubbles.
Yet, they fall short for resolving the imbalances in mini-batches throughout the MLLM training. This gap underpins the motivation behind OrchMLLM, designed to meet the unique challenges of Modality Composition Incoherence.
\section{Conclusion}
In this paper, we introduced OrchMLLM, a comprehensive framework designed to enhance the efficiency and scalability of MLLM training by addressing the issue of Modality Composition Incoherence. We proposed the Batch Post-Balancing Dispatcher and the MLLM Global Orchestrator to mitigate mini-batch imbalances and harmonize multimodal data orchestration. Experimental results demonstrate that OrchMLLM significantly outperforms existing frameworks like Megatron-LM. Hence, OrchMLLM offers a promising solution for efficient and scalable MLLM training, paving the way for future research and development in the field.

\newpage
\bibliographystyle{ACM-Reference-Format}
\bibliography{sample-base}


\begin{thebibliography}{54}


\ifx \showCODEN    \undefined \def \showCODEN     #1{\unskip}     \fi
\ifx \showISBNx    \undefined \def \showISBNx     #1{\unskip}     \fi
\ifx \showISBNxiii \undefined \def \showISBNxiii  #1{\unskip}     \fi
\ifx \showISSN     \undefined \def \showISSN      #1{\unskip}     \fi
\ifx \showLCCN     \undefined \def \showLCCN      #1{\unskip}     \fi
\ifx \shownote     \undefined \def \shownote      #1{#1}          \fi
\ifx \showarticletitle \undefined \def \showarticletitle #1{#1}   \fi
\ifx \showURL      \undefined \def \showURL       {\relax}        \fi
\providecommand\bibfield[2]{#2}
\providecommand\bibinfo[2]{#2}
\providecommand\natexlab[1]{#1}
\providecommand\showeprint[2][]{arXiv:#2}

\bibitem[tre(2020)]%
        {treport}
 \bibinfo{year}{2020}\natexlab{}.
\newblock \bibinfo{title}{OpenAI's GPT-3 Language Model: A Technical Overview}.
\newblock \bibinfo{howpublished}{\url{https://lambdalabs.com/blog/demystifying-gpt-3}}.
\newblock


\bibitem[cha(2022)]%
        {chatgpt}
 \bibinfo{year}{2022}\natexlab{}.
\newblock \bibinfo{title}{Introducing ChatGPT}.
\newblock \bibinfo{howpublished}{\url{https://openai.com/blog/chatgpt}}.
\newblock


\bibitem[gpt(2023)]%
        {gpt4-v}
 \bibinfo{year}{2023}\natexlab{}.
\newblock \bibinfo{title}{GPT-4V(ision) System Card}.
\newblock \bibinfo{howpublished}{\url{https://cdn.openai.com/papers/GPTV_System_Card.pdf}}.
\newblock


\bibitem[gpt(2024)]%
        {gpt4-o}
 \bibinfo{year}{2024}\natexlab{}.
\newblock \bibinfo{title}{Hello GPT-4o}.
\newblock \bibinfo{howpublished}{\url{https://openai.com/index/hello-gpt-4o/}}.
\newblock


\bibitem[gem(2024)]%
        {gemini}
 \bibinfo{year}{2024}\natexlab{}.
\newblock \bibinfo{title}{Introducing Gemini: our largest and most capable AI model}.
\newblock \bibinfo{howpublished}{\url{https://blog.google/technology/ai/google-gemini-ai/}}.
\newblock


\bibitem[tor(2024)]%
        {torchdist}
 \bibinfo{year}{2024}\natexlab{}.
\newblock \bibinfo{title}{PyTorch Distributed Overview}.
\newblock \bibinfo{howpublished}{\url{https://pytorch.org/tutorials/beginner/dist_overview.html}}.
\newblock


\bibitem[Ainslie et~al\mbox{.}(2023)]%
        {gqa}
\bibfield{author}{\bibinfo{person}{Joshua Ainslie}, \bibinfo{person}{James Lee-Thorp}, \bibinfo{person}{Michiel de Jong}, \bibinfo{person}{Yury Zemlyanskiy}, \bibinfo{person}{Federico Lebrón}, {and} \bibinfo{person}{Sumit Sanghai}.} \bibinfo{year}{2023}\natexlab{}.
\newblock \bibinfo{title}{GQA: Training Generalized Multi-Query Transformer Models from Multi-Head Checkpoints}.
\newblock
\showeprint[arxiv]{2305.13245}~[cs.CL]
\urldef\tempurl%
\url{https://arxiv.org/abs/2305.13245}
\showURL{%
\tempurl}


\bibitem[Alayrac et~al\mbox{.}(2022)]%
        {alayrac2022flamingo}
\bibfield{author}{\bibinfo{person}{Jean-Baptiste Alayrac}, \bibinfo{person}{Jeff Donahue}, \bibinfo{person}{Pauline Luc}, \bibinfo{person}{Antoine Miech}, \bibinfo{person}{Iain Barr}, \bibinfo{person}{Yana Hasson}, \bibinfo{person}{Karel Lenc}, \bibinfo{person}{Arthur Mensch}, \bibinfo{person}{Katherine Millican}, \bibinfo{person}{Malcolm Reynolds}, {et~al\mbox{.}}} \bibinfo{year}{2022}\natexlab{}.
\newblock \showarticletitle{Flamingo: a visual language model for few-shot learning}. In \bibinfo{booktitle}{\emph{Advances in Neural Information Processing Systems}}.
\newblock


\bibitem[Athlur et~al\mbox{.}(2022)]%
        {athlur2022varuna}
\bibfield{author}{\bibinfo{person}{Sanjith Athlur}, \bibinfo{person}{Nitika Saran}, \bibinfo{person}{Muthian Sivathanu}, \bibinfo{person}{Ramachandran Ramjee}, {and} \bibinfo{person}{Nipun Kwatra}.} \bibinfo{year}{2022}\natexlab{}.
\newblock \showarticletitle{Varuna: scalable, low-cost training of massive deep learning models}. In \bibinfo{booktitle}{\emph{EuroSys}}.
\newblock


\bibitem[Bai et~al\mbox{.}(2024)]%
        {bai2024seedasrunderstandingdiversespeech}
\bibfield{author}{\bibinfo{person}{Ye Bai}, \bibinfo{person}{Jingping Chen}, \bibinfo{person}{Jitong Chen}, \bibinfo{person}{Wei Chen}, \bibinfo{person}{Zhuo Chen}, \bibinfo{person}{Chuang Ding}, \bibinfo{person}{Linhao Dong}, \bibinfo{person}{Qianqian Dong}, \bibinfo{person}{Yujiao Du}, \bibinfo{person}{Kepan Gao}, \bibinfo{person}{Lu Gao}, \bibinfo{person}{Yi Guo}, \bibinfo{person}{Minglun Han}, \bibinfo{person}{Ting Han}, \bibinfo{person}{Wenchao Hu}, \bibinfo{person}{Xinying Hu}, \bibinfo{person}{Yuxiang Hu}, \bibinfo{person}{Deyu Hua}, \bibinfo{person}{Lu Huang}, \bibinfo{person}{Mingkun Huang}, \bibinfo{person}{Youjia Huang}, \bibinfo{person}{Jishuo Jin}, \bibinfo{person}{Fanliu Kong}, \bibinfo{person}{Zongwei Lan}, \bibinfo{person}{Tianyu Li}, \bibinfo{person}{Xiaoyang Li}, \bibinfo{person}{Zeyang Li}, \bibinfo{person}{Zehua Lin}, \bibinfo{person}{Rui Liu}, \bibinfo{person}{Shouda Liu}, \bibinfo{person}{Lu Lu}, \bibinfo{person}{Yizhou Lu}, \bibinfo{person}{Jingting Ma}, \bibinfo{person}{Shengtao Ma},
  \bibinfo{person}{Yulin Pei}, \bibinfo{person}{Chen Shen}, \bibinfo{person}{Tian Tan}, \bibinfo{person}{Xiaogang Tian}, \bibinfo{person}{Ming Tu}, \bibinfo{person}{Bo Wang}, \bibinfo{person}{Hao Wang}, \bibinfo{person}{Yuping Wang}, \bibinfo{person}{Yuxuan Wang}, \bibinfo{person}{Hanzhang Xia}, \bibinfo{person}{Rui Xia}, \bibinfo{person}{Shuangyi Xie}, \bibinfo{person}{Hongmin Xu}, \bibinfo{person}{Meng Yang}, \bibinfo{person}{Bihong Zhang}, \bibinfo{person}{Jun Zhang}, \bibinfo{person}{Wanyi Zhang}, \bibinfo{person}{Yang Zhang}, \bibinfo{person}{Yawei Zhang}, \bibinfo{person}{Yijie Zheng}, {and} \bibinfo{person}{Ming Zou}.} \bibinfo{year}{2024}\natexlab{}.
\newblock \bibinfo{title}{Seed-ASR: Understanding Diverse Speech and Contexts with LLM-based Speech Recognition}.
\newblock
\showeprint[arxiv]{2407.04675}~[eess.AS]
\urldef\tempurl%
\url{https://arxiv.org/abs/2407.04675}
\showURL{%
\tempurl}


\bibitem[Ben-Nun and Hoefler(2019)]%
        {dp1}
\bibfield{author}{\bibinfo{person}{Tal Ben-Nun} {and} \bibinfo{person}{Torsten Hoefler}.} \bibinfo{year}{2019}\natexlab{}.
\newblock \showarticletitle{Demystifying Parallel and Distributed Deep Learning: An In-depth Concurrency Analysis}.
\newblock \bibinfo{journal}{\emph{ACM Comput. Surv.}} \bibinfo{volume}{52}, \bibinfo{number}{4}, Article \bibinfo{articleno}{65} (\bibinfo{date}{August} \bibinfo{year}{2019}), \bibinfo{numpages}{43}~pages.
\newblock
\showISSN{0360-0300}
\href{https://doi.org/10.1145/3320060}{doi:\nolinkurl{10.1145/3320060}}


\bibitem[Chu et~al\mbox{.}(2024)]%
        {chu2024qwen2audiotechnicalreport}
\bibfield{author}{\bibinfo{person}{Yunfei Chu}, \bibinfo{person}{Jin Xu}, \bibinfo{person}{Qian Yang}, \bibinfo{person}{Haojie Wei}, \bibinfo{person}{Xipin Wei}, \bibinfo{person}{Zhifang Guo}, \bibinfo{person}{Yichong Leng}, \bibinfo{person}{Yuanjun Lv}, \bibinfo{person}{Jinzheng He}, \bibinfo{person}{Junyang Lin}, \bibinfo{person}{Chang Zhou}, {and} \bibinfo{person}{Jingren Zhou}.} \bibinfo{year}{2024}\natexlab{}.
\newblock \bibinfo{title}{Qwen2-Audio Technical Report}.
\newblock
\showeprint[arxiv]{2407.10759}~[eess.AS]
\urldef\tempurl%
\url{https://arxiv.org/abs/2407.10759}
\showURL{%
\tempurl}


\bibitem[Chu et~al\mbox{.}(2023)]%
        {chu2023qwenaudioadvancinguniversalaudio}
\bibfield{author}{\bibinfo{person}{Yunfei Chu}, \bibinfo{person}{Jin Xu}, \bibinfo{person}{Xiaohuan Zhou}, \bibinfo{person}{Qian Yang}, \bibinfo{person}{Shiliang Zhang}, \bibinfo{person}{Zhijie Yan}, \bibinfo{person}{Chang Zhou}, {and} \bibinfo{person}{Jingren Zhou}.} \bibinfo{year}{2023}\natexlab{}.
\newblock \bibinfo{title}{Qwen-Audio: Advancing Universal Audio Understanding via Unified Large-Scale Audio-Language Models}.
\newblock
\showeprint[arxiv]{2311.07919}~[eess.AS]
\urldef\tempurl%
\url{https://arxiv.org/abs/2311.07919}
\showURL{%
\tempurl}


\bibitem[Dai et~al\mbox{.}(2023)]%
        {dai2023instructblip}
\bibfield{author}{\bibinfo{person}{Wenliang Dai}, \bibinfo{person}{Junnan Li}, \bibinfo{person}{Dongxu Li}, \bibinfo{person}{Anthony Tiong}, \bibinfo{person}{Junqi Zhao}, \bibinfo{person}{Weisheng Wang}, \bibinfo{person}{Boyang Li}, \bibinfo{person}{Pascale Fung}, {and} \bibinfo{person}{Steven Hoi}.} \bibinfo{year}{2023}\natexlab{}.
\newblock \showarticletitle{Instruct{BLIP}: Towards General-purpose Vision-Language Models with Instruction Tuning}. In \bibinfo{booktitle}{\emph{Thirty-seventh Conference on Neural Information Processing Systems}}.
\newblock
\urldef\tempurl%
\url{https://openreview.net/forum?id=vvoWPYqZJA}
\showURL{%
\tempurl}


\bibitem[Dean et~al\mbox{.}(2012)]%
        {Dean2012LargeSD}
\bibfield{author}{\bibinfo{person}{Jeffrey Dean}, \bibinfo{person}{Gregory~S. Corrado}, \bibinfo{person}{Rajat Monga}, \bibinfo{person}{Kai Chen}, \bibinfo{person}{Matthieu Devin}, \bibinfo{person}{Quoc~V. Le}, \bibinfo{person}{Mark~Z. Mao}, \bibinfo{person}{Marc'Aurelio Ranzato}, \bibinfo{person}{Andrew~W. Senior}, \bibinfo{person}{Paul~A. Tucker}, \bibinfo{person}{Ke Yang}, {and} \bibinfo{person}{A. Ng}.} \bibinfo{year}{2012}\natexlab{}.
\newblock \showarticletitle{Large Scale Distributed Deep Networks}. In \bibinfo{booktitle}{\emph{Neural Information Processing Systems}}.
\newblock
\urldef\tempurl%
\url{https://api.semanticscholar.org/CorpusID:372467}
\showURL{%
\tempurl}


\bibitem[Dosovitskiy et~al\mbox{.}(2020)]%
        {dosovitskiy2020image}
\bibfield{author}{\bibinfo{person}{Alexey Dosovitskiy}, \bibinfo{person}{Lucas Beyer}, \bibinfo{person}{Alexander Kolesnikov}, \bibinfo{person}{Dirk Weissenborn}, \bibinfo{person}{Xiaohua Zhai}, \bibinfo{person}{Thomas Unterthiner}, \bibinfo{person}{Mostafa Dehghani}, \bibinfo{person}{Matthias Minderer}, \bibinfo{person}{Georg Heigold}, \bibinfo{person}{Sylvain Gelly}, {et~al\mbox{.}}} \bibinfo{year}{2020}\natexlab{}.
\newblock \showarticletitle{An image is worth 16x16 words: Transformers for image recognition at scale}.
\newblock \bibinfo{journal}{\emph{arXiv preprint arXiv:2010.11929}} (\bibinfo{year}{2020}).
\newblock


\bibitem[Feng et~al\mbox{.}(2024)]%
        {feng2024optimus}
\bibfield{author}{\bibinfo{person}{Weiqi Feng}, \bibinfo{person}{Yangrui Chen}, \bibinfo{person}{Shaoyu Wang}, \bibinfo{person}{Yanghua Peng}, \bibinfo{person}{Haibin Lin}, {and} \bibinfo{person}{Minlan Yu}.} \bibinfo{year}{2024}\natexlab{}.
\newblock \showarticletitle{Optimus: Accelerating Large-Scale Multi-Modal LLM Training by Bubble Exploitation}.
\newblock \bibinfo{journal}{\emph{arXiv preprint arXiv:2408.03505}} (\bibinfo{year}{2024}).
\newblock


\bibitem[Hu et~al\mbox{.}(2024)]%
        {hu2024characterization}
\bibfield{author}{\bibinfo{person}{Qinghao Hu}, \bibinfo{person}{Zhisheng Ye}, \bibinfo{person}{Zerui Wang}, \bibinfo{person}{Guoteng Wang}, \bibinfo{person}{Meng Zhang}, \bibinfo{person}{Qiaoling Chen}, \bibinfo{person}{Peng Sun}, \bibinfo{person}{Dahua Lin}, \bibinfo{person}{Xiaolin Wang}, \bibinfo{person}{Yingwei Luo}, {et~al\mbox{.}}} \bibinfo{year}{2024}\natexlab{}.
\newblock \showarticletitle{Characterization of large language model development in the datacenter}. In \bibinfo{booktitle}{\emph{USENIX NSDI}}.
\newblock


\bibitem[Huang et~al\mbox{.}(2024)]%
        {huang2024distmm}
\bibfield{author}{\bibinfo{person}{Jun Huang}, \bibinfo{person}{Zhen Zhang}, \bibinfo{person}{Shuai Zheng}, \bibinfo{person}{Feng Qin}, {and} \bibinfo{person}{Yida Wang}.} \bibinfo{year}{2024}\natexlab{}.
\newblock \showarticletitle{DISTMM: Accelerating Distributed Multimodal Model Training}. In \bibinfo{booktitle}{\emph{USENIX NSDI}}.
\newblock


\bibitem[Jang et~al\mbox{.}(2023)]%
        {jang2023oobleck}
\bibfield{author}{\bibinfo{person}{Insu Jang}, \bibinfo{person}{Zhenning Yang}, \bibinfo{person}{Zhen Zhang}, \bibinfo{person}{Xin Jin}, {and} \bibinfo{person}{Mosharaf Chowdhury}.} \bibinfo{year}{2023}\natexlab{}.
\newblock \showarticletitle{Oobleck: Resilient distributed training of large models using pipeline templates}. In \bibinfo{booktitle}{\emph{ACM SOSP}}.
\newblock


\bibitem[Jeon et~al\mbox{.}(2024)]%
        {jeon2024graphpipe}
\bibfield{author}{\bibinfo{person}{Byungsoo Jeon}, \bibinfo{person}{Mengdi Wu}, \bibinfo{person}{Shiyi Cao}, \bibinfo{person}{Sunghyun Kim}, \bibinfo{person}{Sunghyun Park}, \bibinfo{person}{Neeraj Aggarwal}, \bibinfo{person}{Colin Unger}, \bibinfo{person}{Daiyaan Arfeen}, \bibinfo{person}{Peiyuan Liao}, \bibinfo{person}{Xupeng Miao}, {et~al\mbox{.}}} \bibinfo{year}{2024}\natexlab{}.
\newblock \showarticletitle{Graphpipe: Improving performance and scalability of dnn training with graph pipeline parallelism}.
\newblock \bibinfo{journal}{\emph{arXiv preprint arXiv:2406.17145}} (\bibinfo{year}{2024}).
\newblock


\bibitem[Jiang et~al\mbox{.}(2021)]%
        {sd}
\bibfield{author}{\bibinfo{person}{Juyong Jiang}, \bibinfo{person}{Yingtao Luo}, \bibinfo{person}{Jae~Boum Kim}, \bibinfo{person}{Kai Zhang}, {and} \bibinfo{person}{Sunghun Kim}.} \bibinfo{year}{2021}\natexlab{}.
\newblock \showarticletitle{Sequential Recommendation with Bidirectional Chronological Augmentation of Transformer}.
\newblock \bibinfo{journal}{\emph{CoRR}}  \bibinfo{volume}{abs/2112.06460} (\bibinfo{year}{2021}).
\newblock
\showeprint[arXiv]{2112.06460}
\urldef\tempurl%
\url{https://arxiv.org/abs/2112.06460}
\showURL{%
\tempurl}


\bibitem[Jiang et~al\mbox{.}(2024)]%
        {jiang2024megascale}
\bibfield{author}{\bibinfo{person}{Ziheng Jiang}, \bibinfo{person}{Haibin Lin}, \bibinfo{person}{Yinmin Zhong}, \bibinfo{person}{Qi Huang}, \bibinfo{person}{Yangrui Chen}, \bibinfo{person}{Zhi Zhang}, \bibinfo{person}{Yanghua Peng}, \bibinfo{person}{Xiang Li}, \bibinfo{person}{Cong Xie}, \bibinfo{person}{Shibiao Nong}, {et~al\mbox{.}}} \bibinfo{year}{2024}\natexlab{}.
\newblock \showarticletitle{MegaScale: Scaling large language model training to more than 10,000 GPUs}. In \bibinfo{booktitle}{\emph{USENIX NSDI}}.
\newblock


\bibitem[Li et~al\mbox{.}(2023)]%
        {li2023blip}
\bibfield{author}{\bibinfo{person}{Junnan Li}, \bibinfo{person}{Dongxu Li}, \bibinfo{person}{Silvio Savarese}, {and} \bibinfo{person}{Steven Hoi}.} \bibinfo{year}{2023}\natexlab{}.
\newblock \showarticletitle{Blip-2: Bootstrapping language-image pre-training with frozen image encoders and large language models}.
\newblock \bibinfo{journal}{\emph{arXiv:2301.12597}} (\bibinfo{year}{2023}).
\newblock


\bibitem[Li et~al\mbox{.}(2021)]%
        {li2021sequence}
\bibfield{author}{\bibinfo{person}{Shenggui Li}, \bibinfo{person}{Fuzhao Xue}, \bibinfo{person}{Chaitanya Baranwal}, \bibinfo{person}{Yongbin Li}, {and} \bibinfo{person}{Yang You}.} \bibinfo{year}{2021}\natexlab{}.
\newblock \showarticletitle{Sequence parallelism: Long sequence training from system perspective}.
\newblock \bibinfo{journal}{\emph{arXiv preprint arXiv:2105.13120}} (\bibinfo{year}{2021}).
\newblock


\bibitem[Liu et~al\mbox{.}(2024)]%
        {Liu_2024_CVPR}
\bibfield{author}{\bibinfo{person}{Haotian Liu}, \bibinfo{person}{Chunyuan Li}, \bibinfo{person}{Yuheng Li}, {and} \bibinfo{person}{Yong~Jae Lee}.} \bibinfo{year}{2024}\natexlab{}.
\newblock \showarticletitle{Improved Baselines with Visual Instruction Tuning}. In \bibinfo{booktitle}{\emph{Proceedings of the IEEE/CVF Conference on Computer Vision and Pattern Recognition (CVPR)}}. \bibinfo{pages}{26296--26306}.
\newblock


\bibitem[Liu et~al\mbox{.}(2023a)]%
        {llava}
\bibfield{author}{\bibinfo{person}{Haotian Liu}, \bibinfo{person}{Chunyuan Li}, \bibinfo{person}{Qingyang Wu}, {and} \bibinfo{person}{Yong~Jae Lee}.} \bibinfo{year}{2023}\natexlab{a}.
\newblock \showarticletitle{Visual instruction tuning}.
\newblock \bibinfo{journal}{\emph{arXiv:2304.08485}} (\bibinfo{year}{2023}).
\newblock


\bibitem[Liu et~al\mbox{.}(2023b)]%
        {liu2023janus}
\bibfield{author}{\bibinfo{person}{Juncai Liu}, \bibinfo{person}{Jessie~Hui Wang}, {and} \bibinfo{person}{Yimin Jiang}.} \bibinfo{year}{2023}\natexlab{b}.
\newblock \showarticletitle{Janus: A unified distributed training framework for sparse mixture-of-experts models}. In \bibinfo{booktitle}{\emph{ACM SIGCOMM}}.
\newblock


\bibitem[Panayotov et~al\mbox{.}(2015)]%
        {7178964}
\bibfield{author}{\bibinfo{person}{Vassil Panayotov}, \bibinfo{person}{Guoguo Chen}, \bibinfo{person}{Daniel Povey}, {and} \bibinfo{person}{Sanjeev Khudanpur}.} \bibinfo{year}{2015}\natexlab{}.
\newblock \showarticletitle{Librispeech: An ASR corpus based on public domain audio books}. In \bibinfo{booktitle}{\emph{2015 IEEE International Conference on Acoustics, Speech and Signal Processing (ICASSP)}}. \bibinfo{pages}{5206--5210}.
\newblock
\href{https://doi.org/10.1109/ICASSP.2015.7178964}{doi:\nolinkurl{10.1109/ICASSP.2015.7178964}}


\bibitem[Radford et~al\mbox{.}(2021)]%
        {radford2021learning}
\bibfield{author}{\bibinfo{person}{Alec Radford}, \bibinfo{person}{Jong~Wook Kim}, \bibinfo{person}{Chris Hallacy}, \bibinfo{person}{Aditya Ramesh}, \bibinfo{person}{Gabriel Goh}, \bibinfo{person}{Sandhini Agarwal}, \bibinfo{person}{Girish Sastry}, \bibinfo{person}{Amanda Askell}, \bibinfo{person}{Pamela Mishkin}, \bibinfo{person}{Jack Clark}, {et~al\mbox{.}}} \bibinfo{year}{2021}\natexlab{}.
\newblock \showarticletitle{Learning transferable visual models from natural language supervision}. In \bibinfo{booktitle}{\emph{International conference on machine learning}}.
\newblock


\bibitem[Radford et~al\mbox{.}(2023)]%
        {whisper}
\bibfield{author}{\bibinfo{person}{Alec Radford}, \bibinfo{person}{Jong~Wook Kim}, \bibinfo{person}{Tao Xu}, \bibinfo{person}{Greg Brockman}, \bibinfo{person}{Christine McLeavey}, {and} \bibinfo{person}{Ilya Sutskever}.} \bibinfo{year}{2023}\natexlab{}.
\newblock \showarticletitle{Robust speech recognition via large-scale weak supervision}. In \bibinfo{booktitle}{\emph{Proceedings of the 40th International Conference on Machine Learning}} (Honolulu, Hawaii, USA) \emph{(\bibinfo{series}{ICML'23})}. \bibinfo{publisher}{JMLR.org}, Article \bibinfo{articleno}{1182}, \bibinfo{numpages}{27}~pages.
\newblock


\bibitem[Rajbhandari et~al\mbox{.}(2020)]%
        {rajbhandari2020zero}
\bibfield{author}{\bibinfo{person}{Samyam Rajbhandari}, \bibinfo{person}{Jeff Rasley}, \bibinfo{person}{Olatunji Ruwase}, {and} \bibinfo{person}{Yuxiong He}.} \bibinfo{year}{2020}\natexlab{}.
\newblock \showarticletitle{Zero: Memory optimizations toward training trillion parameter models}. In \bibinfo{booktitle}{\emph{International Conference for High Performance Computing, Networking, Storage and Analysis}}.
\newblock


\bibitem[Rasley et~al\mbox{.}(2020)]%
        {10.1145/3394486.3406703}
\bibfield{author}{\bibinfo{person}{Jeff Rasley}, \bibinfo{person}{Samyam Rajbhandari}, \bibinfo{person}{Olatunji Ruwase}, {and} \bibinfo{person}{Yuxiong He}.} \bibinfo{year}{2020}\natexlab{}.
\newblock \showarticletitle{DeepSpeed: System Optimizations Enable Training Deep Learning Models with Over 100 Billion Parameters}. In \bibinfo{booktitle}{\emph{Proceedings of the 26th ACM SIGKDD International Conference on Knowledge Discovery \& Data Mining}} (Virtual Event, CA, USA) \emph{(\bibinfo{series}{KDD '20})}. \bibinfo{publisher}{Association for Computing Machinery}, \bibinfo{address}{New York, NY, USA}, \bibinfo{pages}{3505–3506}.
\newblock
\showISBNx{9781450379984}
\href{https://doi.org/10.1145/3394486.3406703}{doi:\nolinkurl{10.1145/3394486.3406703}}


\bibitem[Robbins(1951)]%
        {Robbins1951ASA}
\bibfield{author}{\bibinfo{person}{Herbert~E. Robbins}.} \bibinfo{year}{1951}\natexlab{}.
\newblock \showarticletitle{A Stochastic Approximation Method}.
\newblock \bibinfo{journal}{\emph{Annals of Mathematical Statistics}}  \bibinfo{volume}{22} (\bibinfo{year}{1951}), \bibinfo{pages}{400--407}.
\newblock
\urldef\tempurl%
\url{https://api.semanticscholar.org/CorpusID:16945044}
\showURL{%
\tempurl}


\bibitem[Shazeer(2019)]%
        {mqa}
\bibfield{author}{\bibinfo{person}{Noam Shazeer}.} \bibinfo{year}{2019}\natexlab{}.
\newblock \showarticletitle{Fast Transformer Decoding: One Write-Head is All You Need}.
\newblock \bibinfo{journal}{\emph{CoRR}}  \bibinfo{volume}{abs/1911.02150} (\bibinfo{year}{2019}).
\newblock
\showeprint[arXiv]{1911.02150}
\urldef\tempurl%
\url{http://arxiv.org/abs/1911.02150}
\showURL{%
\tempurl}


\bibitem[Shoeybi et~al\mbox{.}(2019)]%
        {shoeybi2019megatron}
\bibfield{author}{\bibinfo{person}{Mohammad Shoeybi}, \bibinfo{person}{Mostofa Patwary}, \bibinfo{person}{Raul Puri}, \bibinfo{person}{Patrick LeGresley}, \bibinfo{person}{Jared Casper}, {and} \bibinfo{person}{Bryan Catanzaro}.} \bibinfo{year}{2019}\natexlab{}.
\newblock \showarticletitle{Megatron-lm: Training multi-billion parameter language models using model parallelism}.
\newblock \bibinfo{journal}{\emph{arXiv preprint arXiv:1909.08053}} (\bibinfo{year}{2019}).
\newblock


\bibitem[Smith et~al\mbox{.}(2022)]%
        {smith2022using}
\bibfield{author}{\bibinfo{person}{Shaden Smith}, \bibinfo{person}{Mostofa Patwary}, \bibinfo{person}{Brandon Norick}, \bibinfo{person}{Patrick LeGresley}, \bibinfo{person}{Samyam Rajbhandari}, \bibinfo{person}{Jared Casper}, \bibinfo{person}{Zhun Liu}, \bibinfo{person}{Shrimai Prabhumoye}, \bibinfo{person}{George Zerveas}, \bibinfo{person}{Vijay Korthikanti}, {et~al\mbox{.}}} \bibinfo{year}{2022}\natexlab{}.
\newblock \showarticletitle{Using deepspeed and megatron to train megatron-turing nlg 530b, a large-scale generative language model}.
\newblock \bibinfo{journal}{\emph{arXiv preprint arXiv:2201.11990}} (\bibinfo{year}{2022}).
\newblock


\bibitem[Su et~al\mbox{.}(2023)]%
        {pandagpt}
\bibfield{author}{\bibinfo{person}{Yixuan Su}, \bibinfo{person}{Tian Lan}, \bibinfo{person}{Huayang Li}, \bibinfo{person}{Jialu Xu}, \bibinfo{person}{Yan Wang}, {and} \bibinfo{person}{Deng Cai}.} \bibinfo{year}{2023}\natexlab{}.
\newblock \showarticletitle{PandaGPT: One Model To Instruction-Follow Them All}.
\newblock \bibinfo{journal}{\emph{arXiv:2305.16355}} (\bibinfo{year}{2023}).
\newblock


\bibitem[Team(2024)]%
        {team2024chameleon}
\bibfield{author}{\bibinfo{person}{Chameleon Team}.} \bibinfo{year}{2024}\natexlab{}.
\newblock \showarticletitle{Chameleon: Mixed-modal early-fusion foundation models}.
\newblock \bibinfo{journal}{\emph{arXiv preprint arXiv:2405.09818}} (\bibinfo{year}{2024}).
\newblock


\bibitem[Thorpe et~al\mbox{.}(2023)]%
        {thorpe2023bamboo}
\bibfield{author}{\bibinfo{person}{John Thorpe}, \bibinfo{person}{Pengzhan Zhao}, \bibinfo{person}{Jonathan Eyolfson}, \bibinfo{person}{Yifan Qiao}, \bibinfo{person}{Zhihao Jia}, \bibinfo{person}{Minjia Zhang}, \bibinfo{person}{Ravi Netravali}, {and} \bibinfo{person}{Guoqing~Harry Xu}.} \bibinfo{year}{2023}\natexlab{}.
\newblock \showarticletitle{Bamboo: Making preemptible instances resilient for affordable training of large DNNs}. In \bibinfo{booktitle}{\emph{USENIX NSDI}}.
\newblock


\bibitem[Vaswani et~al\mbox{.}(2017)]%
        {vaswani2017attention}
\bibfield{author}{\bibinfo{person}{Ashish Vaswani}, \bibinfo{person}{Noam Shazeer}, \bibinfo{person}{Niki Parmar}, \bibinfo{person}{Jakob Uszkoreit}, \bibinfo{person}{Llion Jones}, \bibinfo{person}{Aidan~N Gomez}, \bibinfo{person}{{\L}ukasz Kaiser}, {and} \bibinfo{person}{Illia Polosukhin}.} \bibinfo{year}{2017}\natexlab{}.
\newblock \showarticletitle{Attention is all you need}. In \bibinfo{booktitle}{\emph{Advances in Neural Information Processing Systems}}.
\newblock


\bibitem[Wang et~al\mbox{.}(2024a)]%
        {wang2024qwen2vlenhancingvisionlanguagemodels}
\bibfield{author}{\bibinfo{person}{Peng Wang}, \bibinfo{person}{Shuai Bai}, \bibinfo{person}{Sinan Tan}, \bibinfo{person}{Shijie Wang}, \bibinfo{person}{Zhihao Fan}, \bibinfo{person}{Jinze Bai}, \bibinfo{person}{Keqin Chen}, \bibinfo{person}{Xuejing Liu}, \bibinfo{person}{Jialin Wang}, \bibinfo{person}{Wenbin Ge}, \bibinfo{person}{Yang Fan}, \bibinfo{person}{Kai Dang}, \bibinfo{person}{Mengfei Du}, \bibinfo{person}{Xuancheng Ren}, \bibinfo{person}{Rui Men}, \bibinfo{person}{Dayiheng Liu}, \bibinfo{person}{Chang Zhou}, \bibinfo{person}{Jingren Zhou}, {and} \bibinfo{person}{Junyang Lin}.} \bibinfo{year}{2024}\natexlab{a}.
\newblock \bibinfo{title}{Qwen2-VL: Enhancing Vision-Language Model's Perception of the World at Any Resolution}.
\newblock
\showeprint[arxiv]{2409.12191}~[cs.CV]
\urldef\tempurl%
\url{https://arxiv.org/abs/2409.12191}
\showURL{%
\tempurl}


\bibitem[Wang et~al\mbox{.}(2024b)]%
        {wang2024cogvlmvisualexpertpretrained}
\bibfield{author}{\bibinfo{person}{Weihan Wang}, \bibinfo{person}{Qingsong Lv}, \bibinfo{person}{Wenmeng Yu}, \bibinfo{person}{Wenyi Hong}, \bibinfo{person}{Ji Qi}, \bibinfo{person}{Yan Wang}, \bibinfo{person}{Junhui Ji}, \bibinfo{person}{Zhuoyi Yang}, \bibinfo{person}{Lei Zhao}, \bibinfo{person}{Xixuan Song}, \bibinfo{person}{Jiazheng Xu}, \bibinfo{person}{Bin Xu}, \bibinfo{person}{Juanzi Li}, \bibinfo{person}{Yuxiao Dong}, \bibinfo{person}{Ming Ding}, {and} \bibinfo{person}{Jie Tang}.} \bibinfo{year}{2024}\natexlab{b}.
\newblock \bibinfo{title}{CogVLM: Visual Expert for Pretrained Language Models}.
\newblock
\showeprint[arxiv]{2311.03079}~[cs.CV]
\urldef\tempurl%
\url{https://arxiv.org/abs/2311.03079}
\showURL{%
\tempurl}


\bibitem[Wang et~al\mbox{.}(2024c)]%
        {wang2024cogvlm}
\bibfield{author}{\bibinfo{person}{Weihan Wang}, \bibinfo{person}{Qingsong Lv}, \bibinfo{person}{Wenmeng Yu}, \bibinfo{person}{Wenyi Hong}, \bibinfo{person}{Ji Qi}, \bibinfo{person}{Yan Wang}, \bibinfo{person}{Junhui Ji}, \bibinfo{person}{Zhuoyi Yang}, \bibinfo{person}{Lei Zhao}, \bibinfo{person}{Song XiXuan}, \bibinfo{person}{Jiazheng Xu}, \bibinfo{person}{Xu Bin}, \bibinfo{person}{Juanzi Li}, \bibinfo{person}{Jie Tang}, {and} \bibinfo{person}{Ming Ding}.} \bibinfo{year}{2024}\natexlab{c}.
\newblock \bibinfo{title}{Cog{VLM}: Visual Expert for Large Language Models}.
\newblock
\urldef\tempurl%
\url{https://openreview.net/forum?id=c72vop46KY}
\showURL{%
\tempurl}


\bibitem[Xu et~al\mbox{.}(2025)]%
        {xu2025qwen25omnitechnicalreport}
\bibfield{author}{\bibinfo{person}{Jin Xu}, \bibinfo{person}{Zhifang Guo}, \bibinfo{person}{Jinzheng He}, \bibinfo{person}{Hangrui Hu}, \bibinfo{person}{Ting He}, \bibinfo{person}{Shuai Bai}, \bibinfo{person}{Keqin Chen}, \bibinfo{person}{Jialin Wang}, \bibinfo{person}{Yang Fan}, \bibinfo{person}{Kai Dang}, \bibinfo{person}{Bin Zhang}, \bibinfo{person}{Xiong Wang}, \bibinfo{person}{Yunfei Chu}, {and} \bibinfo{person}{Junyang Lin}.} \bibinfo{year}{2025}\natexlab{}.
\newblock \bibinfo{title}{Qwen2.5-Omni Technical Report}.
\newblock
\showeprint[arxiv]{2503.20215}~[cs.CL]
\urldef\tempurl%
\url{https://arxiv.org/abs/2503.20215}
\showURL{%
\tempurl}


\bibitem[Yang et~al\mbox{.}(2024)]%
        {yang2024qwen2technicalreport}
\bibfield{author}{\bibinfo{person}{An Yang}, \bibinfo{person}{Baosong Yang}, \bibinfo{person}{Binyuan Hui}, \bibinfo{person}{Bo Zheng}, \bibinfo{person}{Bowen Yu}, \bibinfo{person}{Chang Zhou}, \bibinfo{person}{Chengpeng Li}, \bibinfo{person}{Chengyuan Li}, \bibinfo{person}{Dayiheng Liu}, \bibinfo{person}{Fei Huang}, \bibinfo{person}{Guanting Dong}, \bibinfo{person}{Haoran Wei}, \bibinfo{person}{Huan Lin}, \bibinfo{person}{Jialong Tang}, \bibinfo{person}{Jialin Wang}, \bibinfo{person}{Jian Yang}, \bibinfo{person}{Jianhong Tu}, \bibinfo{person}{Jianwei Zhang}, \bibinfo{person}{Jianxin Ma}, \bibinfo{person}{Jianxin Yang}, \bibinfo{person}{Jin Xu}, \bibinfo{person}{Jingren Zhou}, \bibinfo{person}{Jinze Bai}, \bibinfo{person}{Jinzheng He}, \bibinfo{person}{Junyang Lin}, \bibinfo{person}{Kai Dang}, \bibinfo{person}{Keming Lu}, \bibinfo{person}{Keqin Chen}, \bibinfo{person}{Kexin Yang}, \bibinfo{person}{Mei Li}, \bibinfo{person}{Mingfeng Xue}, \bibinfo{person}{Na Ni}, \bibinfo{person}{Pei Zhang},
  \bibinfo{person}{Peng Wang}, \bibinfo{person}{Ru Peng}, \bibinfo{person}{Rui Men}, \bibinfo{person}{Ruize Gao}, \bibinfo{person}{Runji Lin}, \bibinfo{person}{Shijie Wang}, \bibinfo{person}{Shuai Bai}, \bibinfo{person}{Sinan Tan}, \bibinfo{person}{Tianhang Zhu}, \bibinfo{person}{Tianhao Li}, \bibinfo{person}{Tianyu Liu}, \bibinfo{person}{Wenbin Ge}, \bibinfo{person}{Xiaodong Deng}, \bibinfo{person}{Xiaohuan Zhou}, \bibinfo{person}{Xingzhang Ren}, \bibinfo{person}{Xinyu Zhang}, \bibinfo{person}{Xipin Wei}, \bibinfo{person}{Xuancheng Ren}, \bibinfo{person}{Xuejing Liu}, \bibinfo{person}{Yang Fan}, \bibinfo{person}{Yang Yao}, \bibinfo{person}{Yichang Zhang}, \bibinfo{person}{Yu Wan}, \bibinfo{person}{Yunfei Chu}, \bibinfo{person}{Yuqiong Liu}, \bibinfo{person}{Zeyu Cui}, \bibinfo{person}{Zhenru Zhang}, \bibinfo{person}{Zhifang Guo}, {and} \bibinfo{person}{Zhihao Fan}.} \bibinfo{year}{2024}\natexlab{}.
\newblock \bibinfo{title}{Qwen2 Technical Report}.
\newblock
\showeprint[arxiv]{2407.10671}~[cs.CL]
\urldef\tempurl%
\url{https://arxiv.org/abs/2407.10671}
\showURL{%
\tempurl}


\bibitem[Yao et~al\mbox{.}(2024)]%
        {yao2024minicpm}
\bibfield{author}{\bibinfo{person}{Yuan Yao}, \bibinfo{person}{Tianyu Yu}, \bibinfo{person}{Ao Zhang}, \bibinfo{person}{Chongyi Wang}, \bibinfo{person}{Junbo Cui}, \bibinfo{person}{Hongji Zhu}, \bibinfo{person}{Tianchi Cai}, \bibinfo{person}{Haoyu Li}, \bibinfo{person}{Weilin Zhao}, \bibinfo{person}{Zhihui He}, {et~al\mbox{.}}} \bibinfo{year}{2024}\natexlab{}.
\newblock \showarticletitle{MiniCPM-V: A GPT-4V Level MLLM on Your Phone}.
\newblock \bibinfo{journal}{\emph{arXiv preprint arXiv:2408.01800}} (\bibinfo{year}{2024}).
\newblock


\bibitem[Ye et~al\mbox{.}(2022)]%
        {ye2022dbsdynamicbatchsize}
\bibfield{author}{\bibinfo{person}{Qing Ye}, \bibinfo{person}{Yuhao Zhou}, \bibinfo{person}{Mingjia Shi}, \bibinfo{person}{Yanan Sun}, {and} \bibinfo{person}{Jiancheng Lv}.} \bibinfo{year}{2022}\natexlab{}.
\newblock \bibinfo{title}{DBS: Dynamic Batch Size For Distributed Deep Neural Network Training}.
\newblock
\showeprint[arxiv]{2007.11831}~[cs.LG]
\urldef\tempurl%
\url{https://arxiv.org/abs/2007.11831}
\showURL{%
\tempurl}


\bibitem[Yin et~al\mbox{.}(2024)]%
        {yin2024surveymultimodallargelanguage}
\bibfield{author}{\bibinfo{person}{Shukang Yin}, \bibinfo{person}{Chaoyou Fu}, \bibinfo{person}{Sirui Zhao}, \bibinfo{person}{Ke Li}, \bibinfo{person}{Xing Sun}, \bibinfo{person}{Tong Xu}, {and} \bibinfo{person}{Enhong Chen}.} \bibinfo{year}{2024}\natexlab{}.
\newblock \bibinfo{title}{A Survey on Multimodal Large Language Models}.
\newblock
\showeprint[arxiv]{2306.13549}~[cs.CV]
\urldef\tempurl%
\url{https://arxiv.org/abs/2306.13549}
\showURL{%
\tempurl}


\bibitem[Zeng et~al\mbox{.}(2023)]%
        {zeng2023matters}
\bibfield{author}{\bibinfo{person}{Yan Zeng}, \bibinfo{person}{Hanbo Zhang}, \bibinfo{person}{Jiani Zheng}, \bibinfo{person}{Jiangnan Xia}, \bibinfo{person}{Guoqiang Wei}, \bibinfo{person}{Yang Wei}, \bibinfo{person}{Yuchen Zhang}, {and} \bibinfo{person}{Tao Kong}.} \bibinfo{year}{2023}\natexlab{}.
\newblock \showarticletitle{What Matters in Training a GPT4-Style Language Model with Multimodal Inputs?}
\newblock \bibinfo{journal}{\emph{arXiv:2307.02469}} (\bibinfo{year}{2023}).
\newblock


\bibitem[Zhai et~al\mbox{.}(2022)]%
        {zhai2022lit}
\bibfield{author}{\bibinfo{person}{Xiaohua Zhai}, \bibinfo{person}{Xiao Wang}, \bibinfo{person}{Basil Mustafa}, \bibinfo{person}{Andreas Steiner}, \bibinfo{person}{Daniel Keysers}, \bibinfo{person}{Alexander Kolesnikov}, {and} \bibinfo{person}{Lucas Beyer}.} \bibinfo{year}{2022}\natexlab{}.
\newblock \showarticletitle{Lit: Zero-shot transfer with locked-image text tuning}. In \bibinfo{booktitle}{\emph{IEEE Conference on Computer Vision and Pattern Recognition}}.
\newblock


\bibitem[Zhang et~al\mbox{.}(2023)]%
        {pmc-vqa}
\bibfield{author}{\bibinfo{person}{Xiaoman Zhang}, \bibinfo{person}{Chaoyi Wu}, \bibinfo{person}{Ziheng Zhao}, \bibinfo{person}{Weixiong Lin}, \bibinfo{person}{Ya Zhang}, \bibinfo{person}{Yanfeng Wang}, {and} \bibinfo{person}{Weidi Xie}.} \bibinfo{year}{2023}\natexlab{}.
\newblock \showarticletitle{PMC-VQA: Visual Instruction Tuning for Medical Visual Question Answering}.
\newblock \bibinfo{journal}{\emph{arXiv:2305.10415}} (\bibinfo{year}{2023}).
\newblock


\bibitem[Zhang et~al\mbox{.}(2024)]%
        {zhang2024disttrainaddressingmodeldata}
\bibfield{author}{\bibinfo{person}{Zili Zhang}, \bibinfo{person}{Yinmin Zhong}, \bibinfo{person}{Ranchen Ming}, \bibinfo{person}{Hanpeng Hu}, \bibinfo{person}{Jianjian Sun}, \bibinfo{person}{Zheng Ge}, \bibinfo{person}{Yibo Zhu}, {and} \bibinfo{person}{Xin Jin}.} \bibinfo{year}{2024}\natexlab{}.
\newblock \bibinfo{title}{DistTrain: Addressing Model and Data Heterogeneity with Disaggregated Training for Multimodal Large Language Models}.
\newblock
\showeprint[arxiv]{2408.04275}~[cs.DC]
\urldef\tempurl%
\url{https://arxiv.org/abs/2408.04275}
\showURL{%
\tempurl}


\bibitem[Zhao et~al\mbox{.}(2023)]%
        {zhao2023pytorch}
\bibfield{author}{\bibinfo{person}{Yanli Zhao}, \bibinfo{person}{Andrew Gu}, \bibinfo{person}{Rohan Varma}, \bibinfo{person}{Liang Luo}, \bibinfo{person}{Chien-Chin Huang}, \bibinfo{person}{Min Xu}, \bibinfo{person}{Less Wright}, \bibinfo{person}{Hamid Shojanazeri}, \bibinfo{person}{Myle Ott}, \bibinfo{person}{Sam Shleifer}, {et~al\mbox{.}}} \bibinfo{year}{2023}\natexlab{}.
\newblock \showarticletitle{Pytorch fsdp: experiences on scaling fully sharded data parallel}.
\newblock \bibinfo{journal}{\emph{arXiv preprint arXiv:2304.11277}} (\bibinfo{year}{2023}).
\newblock


\end{thebibliography}

\newpage

\appendix

\section{Other Post-Balancing Algorithms}
\label{app:other_alg}

\begin{algorithm}[h]
\caption{Post-Balancing Algorithm 3rd}\label{alg:rmpad_squ}
    \begin{footnotesize}
    \begin{algorithmic}[1]
        \Require count of DP instances $d$, list of sequences $S$, tolerance interval $v$ 
        \Function{CMP}{a, b}
        \If{$\text{abs}(a.\text{lengths\_sum()}-b.\text{lengths\_sum()})<v$}
        \State \Return $a.\text{lengths\_square\_sum()}<b.\text{lengths\_square\_sum()}$
        \EndIf
        \State \Return $a.\text{lengths\_sum()}<b.\text{lengths\_sum()}$
        \EndFunction
        \State $\mathit{sorted\_sequences} \gets$ Sort $S$ in descending order by length, 
        \State Initialize $\mathit{new\_batches}$ as a priority queue that sort the batches with comparative function \Call{CMP}{}, 
        \For{$i=1 \rightarrow d$}
        \State $B_i \gets \emptyset$, $\mathit{new\_batches.\text{push}(B_i)}$
        \EndFor
        \For{$s \in sorted\_sequences$}
        \State $\mathit{new\_batches.\text{top}().\text{push}(s)}$
        \EndFor
        \State \Return $\mathit{new\_batches.\text{tolist}()}$
    \end{algorithmic}
    \end{footnotesize}
\end{algorithm}

Then, we discuss the scenario where the assumption $\beta \ll \alpha$ is not valid (still in use of the classic transformer architecture). In this part, we only consider the batching method without paddings, hence the objective is given by:
\begin{equation*}
 \text{Objective:} \quad  \underset{\Pi}{\text{minimize}} ~ \max_{0\leq i < d} L_i'(\Pi) + \lambda\sum_{j=0}^{b_i - 1} (l'_{i,j}(\Pi))^2.
\end{equation*}
where $\lambda=\frac{\beta}{\alpha}.$ The approximation algorithm is presented as Algorithm~\ref{alg:rmpad_squ}. The computational complexity is $O(n \log n)$, the same as that of Algorithm~\ref{alg:rmpad}. The tolerance interval is manually set to trade off between linear and quadratic terms.

\begin{algorithm}[h!]
\caption{Post-Balancing Algorithm 4th}\label{alg:rmpad_pad}
    \begin{footnotesize}
    \begin{algorithmic}[1]
    
        \Require count of DP instances $d$, list of sequences $S$
        \State $max\_sum\_bounds \gets $ the objective value of Post-Balancing Algorithm without Paddings,
        \State $\mathit{sorted\_sequences} \gets $ Sort $S$ in descending order by length,
        \State $\mathit{new\_batches} \gets \{\{\}\}$
        \For{$s \in sorted\_sequences$}
        \If{$(\text{len}(\mathit{new\_batches}[-1]) + 1) * s.\text{length} > b$}
        \If{$\text{len} > d$}
        \State break
        \EndIf
        \State $\mathit{new\_batches}.\text{push}(\{\})$
        \EndIf
        \State $\mathit{new\_batches}[-1].\text{push}(s)$
        \EndFor
        \State $sorted_sequences \gets $ remaining sequences in $sorted_sequences$,
        \State Transform $new\_batches$ into a priority queue that sort the batches based on the sum of sequence lengths,
        \For{$s \in sorted\_sequences$}
        \State $\mathit{new\_batches.\text{top}().\text{push}(s)}$
        \EndFor
        \State \Return $\mathit{new\_batches.\text{tolist}()}$

    \end{algorithmic}
    \end{footnotesize}
\end{algorithm}
Our method is also applicable to non-classical transformer architectures. For example, the architecture of ConvTransformer is is sometimes used for feature extraction in images and speech. The main difference from the transformer lies in the structure of the Attention mechanism. Therefore, it requires padding for computation during the attention phase, rather than using the flash attention operator. The objective for the Post-Balancing problem is given by:
\begin{equation*}
 \text{Objective:} \quad  \underset{\Pi}{\text{minimize}} ~ \max_{0\leq i < d} L_i'(\Pi) + \lambda b_i (\max_{0 \leq j<b_i} l'_{i,j}(\Pi))^2.
\end{equation*}
The approximation algorithm is presented as Algorithm~\ref{alg:rmpad_pad}. The computational complexity is also $O(n \log n)$.

\section{Communication Latency Deduction}

\label{app:deduct}
\medskip\noindent\textbf{Equation~\ref{equ:ag}.}
We deduce the communication overhead of the All-Gather operation with the ring-
based algorithm deployed. In an All-Gather operation, $(d-1)$ circles of communication is needed because of the ring-based communication topology, where $d$ denotes the total number of instances. Since the volume of data communicated is directly proportional to the size of the mini-batch, with the coefficient denoted as $k$, the largest mini-batch size among all instances, $\max_{0 \leq i < d} (L_i)$, dominates the communication cost of each circle. Moreover, the communication bandwidth $B$ of this operation is determined by the lowest bandwidth between two instances in the circle. Therefore, the overall communication overhead incurred by the All-Gather operation can be expressed as $O_\text{All-Gather} = \frac{k}{B}(d-1) \max_{0 \leq i < d} (L_i)$. Therefore, there is a proportional relationship as shown in Equation~\eqref{equ:ag}.

\medskip\noindent\textbf{Equation~\ref{equ:ata}.}
The communication overhead of the All-to-All operation arises from its point-to-point communication protocol, where each instance in the distributed system exchanges data directly with every other instance. Unlike All-Gather, where all data must be shared globally, All-to-All distributes specific portions of the data such that each instance sends and receives targeted mini-batch examples according to the rearrangement requirements. The communication volume between certain pairs of instances is difficult to express, but the maximum of the volume must be restricted under the largest mini-batch size among all instances, $k*\max_{0 \leq i < d} (L_i)$, where it is impossible to meet this upper bound due to the objective of the Post-Balancing algorithms. 
Meanwhile, the lower bound of the communication bandwidth $B_{min}$ of this operation is determined by the lowest bandwidth between all pairs of instances in this system.
Given that the communication process involves individual exchanges between pairs of instances, the communication overhead is bounded by an upper limit $\Omega_\text{All-to-All} = \frac{k}{B_{min}} \max_{0 \leq i < d} (L_i)$. Therefore, the relationship is given as Equation~\eqref{equ:ata}, because the upper limit is impossible to reach. This formulation accounts for the reduced complexity compared to collective broadcast patterns like All-Gather, as the data exchange is distributed more efficiently across the instances based on the point-to-point protocol.

\medskip\noindent\textbf{Equation~\ref{equ:node}.}
In large-scale distributed training clusters, the communication overhead of the All-to-All operation is influenced by the hierarchical communication topologies, where intra-node communication (i.e., communication between instances on the same physical node) often significantly outperforms inter-node communication (i.e., communication between instances spanning different nodes). Typically, intra-node communication leverages high-speed interconnects (e.g., NVLink), with hundreds of GBs of point-to-point bandwidth. 
Conversely, inter-node communication using interconnects like InfiniBand or Ethernet is constrained by significantly lower available bandwidths compared to intra-node communication. This limitation arises because, in many cluster configurations, there are no direct point-to-point connections established between instances on different nodes. Instead, the inter-node communication relies on shared resources such as the InfiniBand network fabric, leading to bandwidth contention. During the All-to-All operation, instances on the same node must compete for the node's allocated InfiniBand bandwidth when communicating with remote instances located on other nodes. As a result, the effective bandwidth available to each instance for inter-node communication is significantly reduced, with the average InfiniBand bandwidth per instance being lower compared to the theoretically available bandwidth of the entire interconnect. 

This disparity leads to stragglers during the All-to-All communication, as inter-node data exchanges become a bottleneck. Under the point-to-point protocol, the communication volume between specific pairs of instances is proportional to $l_{i,j}$—the portion of data exchanged between instance $i$ and instance $j$ after the balancing rearrangement $\Pi$. For each instance $i$, the total inter-node communication is determined by the data volume of whose $i'(\Pi)$ is not residing on the same node of $i$. Thus, assuming that each instance is allocated with the same inter-node communication bandwidth $B_{inter}$ (to simplify the analysis), the maximum inter-node communication of an instance in the system, given by $k*\max_{0 \leq i < d} (\sum_{i'(\Pi) \notin N(i)} l_{i,j})$, will dominates the overall communication overhead. Therefore, the communication latency is denoted as $\frac{k}{B_{inter}}\max_{0 \leq i < d} (\sum_{i'(\Pi) \notin N(i)} l_{i,j})$ and leads to the proportional relationship as shown in Equation~\eqref{equ:node}. This equation highlights that inter-node communication pairs are the primary contributors to the communication overhead, further exacerbating inefficiencies in heterogeneous clusters.

\end{document}